\documentclass[conference]{IEEEtran}

\usepackage{threeparttable} 
\usepackage{multirow}  
\usepackage{makecell} 
\usepackage{arydshln}  

\usepackage{amssymb} 
\usepackage{amsmath} 
\usepackage{xcolor} 
\usepackage{pifont} 
\usepackage{fontawesome5}  
\usepackage{wasysym}  
\usepackage{gensymb}  
\usepackage[symbol]{footmisc}
\usepackage{marvosym} 

\usepackage{rotating} 
\usepackage{balance}  
\usepackage[inline]{enumitem}  
\usepackage{url}
\usepackage{array}

\usepackage[font=small, labelsep=period]{caption}
\usepackage{subcaption}  
\usepackage{graphicx} 

\usepackage[numbers,sort&compress]{natbib} 
\usepackage[colorlinks=true, linkcolor=black, citecolor=black, urlcolor=blue]{hyperref} 

\begin{document}
%

\title{SoK: Understanding the Fundamentals and Implications of Sensor Out-of-band Vulnerabilities}

\author{
        \IEEEauthorblockN{Shilin Xiao, Wenjun Zhu, Yan Jiang, Kai Wang, Peiwang Wang, Chen Yan, Xiaoyu Ji\textsuperscript{*}, Wenyuan Xu}
        \IEEEauthorblockA{Zhejiang University\\
        \{xshilin, zwj\_, yj98, eekaiwang, wangpw, yanchen, xji, wyxu\}@zju.edu.cn}
        }


%


\IEEEoverridecommandlockouts
\makeatletter\def\@IEEEpubidpullup{4\baselineskip}\makeatother
\IEEEpubid{\parbox{\columnwidth}{
		Network and Distributed System Security (NDSS) Symposium 2026\\
		23-27 February 2026, San Diego, CA, USA\\
		ISBN 979-8-9894372-8-3\\
		https://dx.doi.org/10.14722/ndss.2026.230450\\
		www.ndss-symposium.org
}
\hspace{\columnsep}\makebox[\columnwidth]{}}

\maketitle
\footnotetext{\textsuperscript{*} Xiaoyu Ji is the corresponding author.}


\begin{abstract}

Sensors are fundamental to cyber-physical systems (CPS), enabling perception and control by transducing physical stimuli into digital measurements. However, despite growing research on physical attacks on sensors, our understanding of sensor hardware vulnerabilities remains fragmented due to the ad-hoc nature of this field. Moreover, the infinite attack signal space further complicates threat abstraction and defense. To address this gap, we propose a systematization framework, termed sensor \textit{out-of-band} (OOB) vulnerabilities, that for the first time provides a comprehensive abstraction for sensor attack surfaces based on underlying physical principles. 
We adopt a bottom-up systematization methodology that analyzes OOB vulnerabilities across three levels. At the component level, we identify the physical principles and limitations that contribute to OOB vulnerabilities. At the sensor level, we categorize known attacks and evaluate their practicality. At the system level, we analyze how CPS features such as sensor fusion, closed-loop control, and intelligent perception impact the exposure and mitigation of OOB threats. Our findings offer a foundational understanding of sensor hardware security and provide guidance and future directions for sensor designers, security researchers, and system developers aiming to build more secure sensors and CPS.

\end{abstract}
\section{Introduction}

Sensors serve as the essential bridge between the physical and cyber worlds by transducing physical stimuli into digital signals. They are widely deployed in cyber-physical systems (CPS), ranging from industrial robots to critical infrastructure, and are crucial in providing correct measurements for decision-making in safety-sensitive CPS. Incorrect sensor measurements have been linked to major safety incidents, including factory explosions~\cite{aiche2018process}, airplane crashes~\cite{cnn2019boeing}, and fatalities caused by industrial robots~\cite{cbs2023robot}. These events highlight the importance of ensuring sensor reliability as a fundamental requirement for CPS security.

Unfortunately, sensors are vulnerable to physical-world attacks, such as signal injection attacks and side-channel attacks. In signal injection attacks, sensor measurements can be manipulated by various physical signals~\cite{zhang2017dolphinattack,trippel2017walnut,michalevsky2014gyrophone,wang2023volttack}. For instance, ultrasound or lasers modulated with voices can inject malicious commands into microphones, while carefully-crafted electromagnetic signals can induce fake touchpoints on touchscreens. In side-channel attacks, sensors can unintentionally leak sensitive information through TEMPEST-like side channels~\cite{ni2023recovering,cronin2021charger,shi2021face}, where electromagnetic radiation can be exploited to extract biometric data such as fingerprints or iris patterns. These works reveal that sensors exhibit non-negligible hardware vulnerabilities that have become severe threat vectors in CPS.

However, despite extensive research uncovering various threat vectors, our understanding of sensor vulnerabilities remains limited, due to the standalone and ad-hoc nature of this research area. Furthermore, the virtually infinite combinations of physical signals that can compromise sensor security make it difficult to abstract these threats comprehensively. As a result, a unified and effective framework that captures both known and potential vulnerabilities is urgent for improving sensor security but has yet to be fully established. While several SoK papers and surveys have examined sensor security, they mainly focus on unifying attack methodologies~\cite{giechaskiel2019taxonomy, yan2020sok} or analyzing specific scenarios~\cite{sikder2021survey, walker2021sok, xu2023sok}, without providing a profound understanding of sensor vulnerabilities themselves. This lack of systematic understanding continues to hinder the effective detection and defense strategies.

\begin{figure*}[t]
    \centering
    \begin{subfigure}[b]{0.98\columnwidth}
        \centering
        \includegraphics[width=0.98\textwidth]{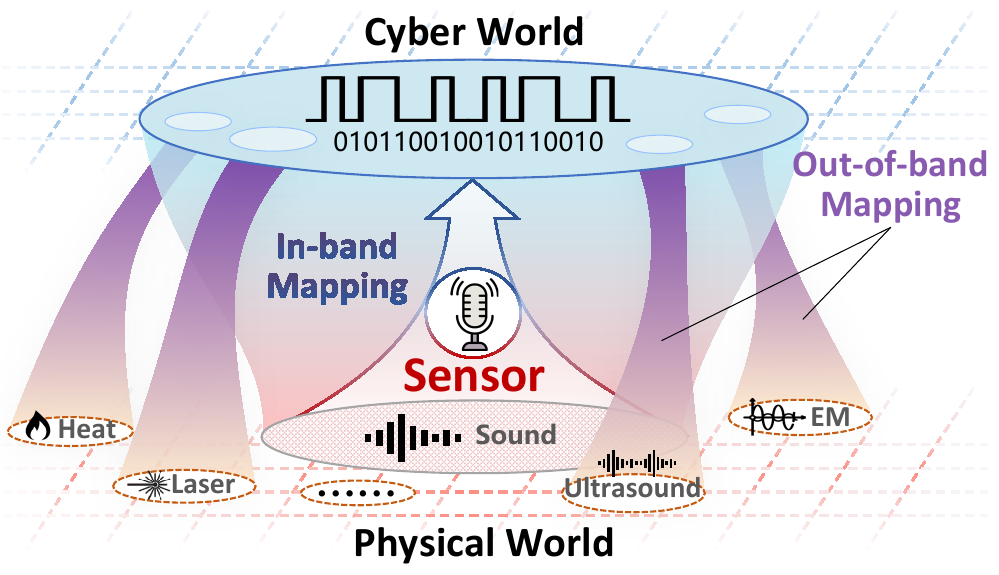}
        \caption{In-band mapping v.s. Out-of-band mapping}
        \label{fig:in_vs_oob}
    \end{subfigure}
    \begin{subfigure}[b]{0.98\columnwidth}
        \centering
        \includegraphics[width=0.98\textwidth]{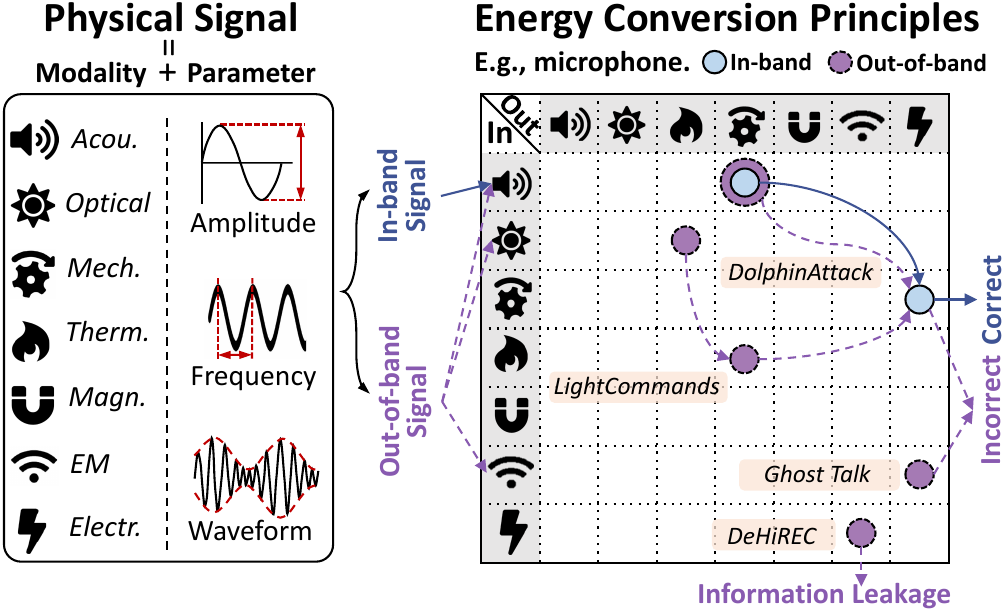}
        \caption{Energy conversion principles}
        \label{fig:energy_onversion}
    \end{subfigure}
    \caption{Illustration of sensor \textit{in-band} and \textit{out-of-band} mappings. \textit{In-band} reflects a sensor's intended functionality, while out-of-band refers to unintended functionality between physical and cyber worlds.}
    \label{fig:OOB}
\end{figure*}

To bridge this gap, this paper integrates previous research and domain knowledge with physical principles to develop a systematic framework for categorizing and analyzing threat vectors.
We begin by examining the fundamental operation of sensors. Each sensor establishes a \textit{mapping} between physical stimuli and digital measurements, which we refer to as \textit{in-band mapping}. For example, the in-band mapping of a microphone converts audible sounds into voice recordings. However, studies have shown that lasers or ultrasound can also produce voice recordings, as illustrated in Fig.~\ref{fig:in_vs_oob}. These interactions fall outside the sensor's intended operational scope and are referred to as \textit{out-of-band} (OOB) mappings. OOB mappings stem from the inherent limitations of sensor components, such as material, mechanical, electrical non-idealities. Accordingly, we define sensor \textit{out-of-band} vulnerability as the security risk caused by OOB mappings.

To abstract OOB threat vectors, we model attack signals in terms of their modality and physical parameters, as illustrated in Fig.~\ref{fig:energy_onversion}. From a physical perspective, measurand modalities can be classified into seven categories: acoustic, optical, mechanical, thermal, magnetic, electromagnetic, and electrical~\cite{fraden2004handbook}. Each modality can be further characterized by its amplitude, frequency, and waveform.
The fundamental interactions between attack signals and sensor OOB vulnerabilities can be captured using a $7\times7$ matrix, where each cell represents a specific energy conversion pathway, e.g., mechanical-to-electrical transduction. A vulnerability exists if there is a viable path through this matrix that starts with an injected signal modality and ends in the electrical column, ultimately producing a measurable digital output.
For instance, the LightCommands attack~\cite{sugawara2020light} exploits a chain of conversions: optical-to-thermal, thermal-to-mechanical, and mechanical-to-electrical. Similarly, EM side-channel leakage, as demonstrated by DeHiREC~\cite{zhou2023dehirec}, leverages an electrical-to-electromagnetic conversion pathway, as shown in Fig.~\ref{fig:energy_onversion}.
Building on this abstraction, we further analyze the underlying mechanisms of OOB vulnerabilities across key sensor components (e.g., amplifiers, filters, transducers), and identify threat vectors that may emerge during different stages of sensor design, thereby providing guidance for  secure sensor design practices.

Based on the modeled energy conversion pathways, we classify OOB vulnerabilities into two types: \textit{out-of-range} and \textit{cross-field}.
\textit{Out-of-range} vulnerabilities arise when the attack signal shares the same physical modality as the intended signal but exceeds the sensor’s design limits in amplitude or frequency. In contrast, \textit{cross-field} vulnerabilities involve signals from unintended physical modalities that exploit unintended energy conversions to manipulate sensor measurements or induce electrical signals through side channels.

While many attacks have been demonstrated in laboratory settings, some studies suggest their practicality may be limited in real-world environments~\cite{walker2021sok}, or that they pose minimal threat to specific CPS applications~\cite{kim2024systematic}. To assess real-world impact, we evaluate each attack across multiple dimensions: attacker prior knowledge required, effective attack distance, attack cost that includes device cost, size. This evaluation gain insights to guide future directions for improving exploitation efficiency and uncovering new attack surfaces. Furthermore, we systematically show that none of the key CPS characteristics, such as closed-loop control, multi-sensor fusion, and intelligent perception, can completely defend against OOB vulnerabilities, and highlight their respective strengths and limitations against sensor attacks. This analysis helps system developers prioritize and tailor their defense strategies.

We summarize our contributions as follows. 
\begin{itemize}[leftmargin=10pt]  
    \item To the best of our knowledge, we proposed the first systematic framework that provides a comprehensive abstraction for sensor threats at the signal level, which essentially models sensor out-of-band (OOB) vulnerabilities and elucidate their core mechanisms, providing a theoretical framework for efficient vulnerability detection.  
    \item We model and analyze existing OOB exploitation methods to evaluate their feasibility in practical settings and identify critical research gaps, utilizing the proposed systematic framework and validating its generalizability.
    \item We explore the system-level implications of OOB vulnerabilities, revealing how different CPS architectures react to sensor attacks, thus offering enhanced security guidance for system designers.  
\end{itemize}

\section{Background and Threat model} 

In this section, we introduce the background of sensor hardware composition, identify the threat model of sensor OOB vulnerability, and clarify our research scope.

\subsection{Sensor Background}\label{sec:background}
\begin{figure}[t!]
    \centering
    \includegraphics[width=0.98\linewidth]{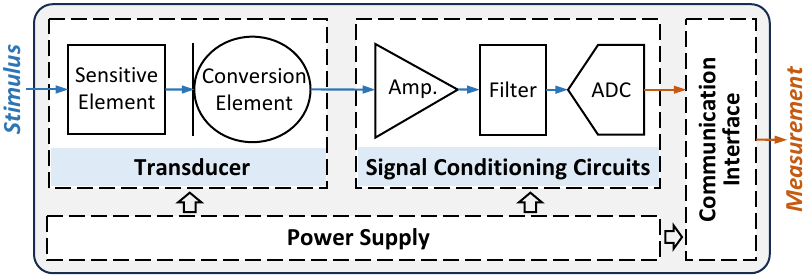}
    \caption{The hardware composition of a modern sensor. }
    \label{fig:sensor}
\end{figure}

A sensor is designed to respond to a physical stimulus and generate electrical measurements. 
Sensors can be classified into over 350 categories based on their measurand types \cite{listofsensor} and vary widely in function and operational principles. Nevertheless, a modern sensor comprises four main components: transducer, signal conditioning circuits, communication interface, and power supply~\cite{webster2018measurement}, as shown in Fig.~\ref{fig:sensor}.

\begin{itemize}[leftmargin=10pt]
    \item \underline{\textit{Transducer.}} Transducers respond to physical stimuli and consist of sensitive and conversion elements~\cite{yu2021different}. The former detects a physical stimulus, and the latter transforms the physical stimulus into an analog electrical signal. 
  
    \item \underline{\textit{Signal conditioning circuits.}} Since the analog signal from the transducing process is noisy and weak, the signal conditioning circuits have to amplify it, reduce noise, digitize it, and even perform additional digital signal processing before measurement can be utilized. These are typically accomplished by amplifiers, filters, ADCs, etc.

    \item \underline{\textit{Communication interface.}} The communication interface enables sensor measurements to be transferred for calculation or decision-making and contains interface circuits with a chosen protocol, e.g., I2C or SPI. 

    \item \underline{\textit{Power supply.}} Typically, a DC power supply is adopted to provide energy to ensure the continuous operation of other sensor components, and may include a transformer, rectifier, filter, and regulator that are designed to control and modulate the power supply. 
\end{itemize}

Each component may introduce sensor vulnerabilities, and we will analyze their root cause and implications at the system level in the following section. Note that Fig.~\ref{fig:sensor} represents a typical composition of most sensors, but special sensors may include extra components, e.g., an active sensor such as LiDARs may incorporate signal transmitters.

\subsection{Threat Model}\label{sec:attack_goals}

We identify a common threat model from existing work. 
\begin{itemize} [leftmargin=10pt]
    \item \underline{\textit{Attacker capabilities.}} 
        \textbf{a) Knowledge.} A conservative assumption is that attackers have no prior knowledge of the target sensors and their associated systems. In practice, most studies assume attackers can gain partial knowledge by prestudying identical devices or reading related documentation. 
        \textbf{b) Accessibility to victim sensors.} Most studies assume that attackers have no physical access to victim's sensors to avoid raising the victim's awareness. Instead, attackers can determine whether an attack succeeds by indirect feedback from the system, such as changes in LED indicators or system behavior. In some scenarios, attackers can also implant malware on the victim's device to access sensor readings.
        \textbf{c) Attack devices.} Attackers can use commonly available devices such as signal generators, amplifiers, speakers, laser emitters, and antennas to inject signals or capture side-channel emissions. Moreover, they can customize attack devices to meet requirements for portability and stealth.
    \item \underline{\textit{Attack goals.}} The attack goals can be classified into three types: 
        \textbf{a) Denial-of-Service (DoS).} The attacker aims to make the measurement unavailable by overwhelming it with high-intensity noise, such as emitting ultrasound to jam ultrasonic sensors~\cite{yan2016can}. 
        \textbf{b) Measurement spoofing.} The attacker spoofs the sensor to produce seemingly legitimate but erroneous measurements, such as creating fake touchpoints on touchscreens by injecting EMI signals into the power cable~\cite{jiang2022wight}. 
        \textbf{c) Privacy Snooping.} The attacker can exploit the side-channel leakage of a sensor to recover private information, such as inferring the victim's keystrokes by recognizing the touchscreen's electromagnetic emanations~\cite{jin2021periscope}.
    \item \underline{\textit{Defense goals.}} We adapt the well-known CIA (Confidentiality, Integrity, and Availability) triad specifically to the context of sensor security. While achieving a perfect CIA triad is nearly impossible, these goals provide general guidelines to help designers understand and prioritize the defenses against sensor attacks.
        \textbf{a) Availability.} A sensor's functionality shall be available to the system in spite of disturbances from the outside world.
        \textbf{b) Integrity.} A sensor's sensitive measurement shall correctly reflect the physical quantity being measured according to its design function.
        \textbf{c) Confidentiality.} A sensor's measurement shall not be divulged to the outside world in the form of any physical signals.
    Note that the three defense goals align with the aforementioned three attack goals: (1) \textit{availability} $\leftrightarrow$ \textit{denial-of-service}, (2) \textit{integrity} $\leftrightarrow$ \textit{spoofing}, and (3) \textit{confidentiality} $\leftrightarrow$ \textit{snooping}.
\end{itemize}

\subsection{Scope of the Study}

To provide a more focused systematization and clarify our research scope, the following topics are excluded. \textbf{a) Sensors for security purpose and atypical sensors.} This excludes sensors that are specifically used to detect tampering or signal injection attacks in CPS, and atypical sensors without transducers, such as time-to-digital converters and ring oscillators.
\textbf{b) Cyber-domain security on sensor-involved systems.} This excludes sensor network security research~\cite{padmavathi2009survey} and physical adversarial example attacks~\cite{kurakin2018adversarial}, which do not exploit sensor vulnerabilities.
\textbf{c) Privacy snooping using sensor's intended functionality.} This excludes attacks such as eavesdropping via radar~\cite{sami2020lidarphone, hu2023mmecho}, keystroke inference via motion sensors~\cite{xu2012taplogger, liu2015good, maiti2016smartwatch, miluzzo2012tapprints, owusu2012accessory} or microphones~\cite{berger2006dictionary, foo2010timing, liu2015snooping, lu2019keylistener, shumailov2019hearing, zhu2014context}, and identity inference via motion sensors~\cite{shi2021face}.

\section{Component-Level Vulnerability Analysis} \label{sec:component-level}

In this section, we present a component-level analysis of sensor OOB vulnerabilities.
Specifically, we formalize an OOB model that categorizes OOB vulnerabilities into two types: \textit{out-of-range} and \textit{cross-field}, and thoroughly analyze their underlying principles. Based on this, we summarize the exploited and potential OOB vulnerabilities across different sensor components, which can further guide the secure sensor design and help prioritize testing strategies.

\subsection{Out-of-band Vulnerability Model} \label{sec:vuln_model}
Sensor \textit{out-of-band} (OOB) vulnerabilities refer to security flaws resulting from unintended mappings between physical-world signals and cyber-world measurements, which is beyond the sensor's intended functionality.
The OOB vulnerability model is presented in Fig.~\ref{fig:vuln_model}.
The intended functionality of each sensor component is formalized as an input-output mapping function $y=f(x)$, i.e., \textit{in-band mapping}.
However, due to non-ideal factors such as material limitations and electrical properties, sensor components may exhibit unintended behaviors, namely \textit{out-of-band mapping}. Specifically, these unintended mappings can be categorized into \textit{out-of-range} mappings and \textit{cross-field} mappings, described as follows:

\begin{figure}[t!]
    \centering
    \includegraphics[width=\linewidth]{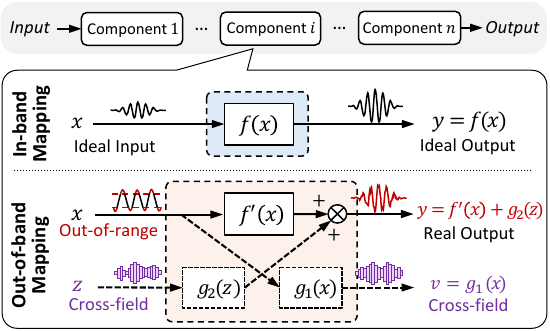}
    \caption{Sensor OOB vulnerability model. We use the input-output mapping functions to formalize the intended and unintended behaviors of sensors.}
    \label{fig:vuln_model}
\end{figure}

\begin{itemize}[leftmargin=15pt]
    \item \underline{\textit{Out-of-range mapping} $f'(x)$} represents the response to an input signal $x$ that lies within the intended physical field, but falls outside the intended operational range, including amplitude and frequency. 
    For instance, the acoustic transducer of a microphone is intended to receive audible sounds, ranging from 20 Hz to 20 kHz, but it can also respond to ultrasonic signals ($>$20 kHz) that are out of the intended frequency range, which enables malicious inaudible voice command injection~\cite{zhang2017dolphinattack}.
    \item \underline{\textit{Cross-field mapping} $g(\cdot)$} refers to the unintended signal interactions across different physical fields and consists of two directions: 
    \textbf{a) cross-field output} $g_1(x)$ describes the case where an internal signal $x$ unintentionally radiates or leaks signals $v$ into another physical field (e.g., optical, acoustic, thermal, or electromagnetic). For example, circuit wires may inadvertently act as antennas, emitting electromagnetic interference (EMI) that leaks information about internal signals~\cite{esteves2018remote}.
    \textbf{b) cross-field input} $g_2(z)$ models the reverse situation, where the system receives an input $z$ from a non-intended physical field, leading to unintended outputs $g_2(z)$ that superimpose with or distort the intended output $y$. For example, MEMS microphones designed to pick up sound can also be stimulated by modulated laser light, leading to injected signals or commands~\cite{sugawara2020light}.
\end{itemize}

\subsection{Fundamentals of Out-of-band Vulnerability} \label{sec:mechanisms}

To further investigate the fundamentals of sensor OOB vulnerabilities, we systematize the vulnerabilities of various sensor components and their mechanisms, as summarized in Table~\ref{tab:vuln_tax}. Specifically, we identify which components contribute to either \textit{out-of-range} vulnerability or \textit{cross-field} vulnerability.

\renewcommand{\arraystretch}{1.3}
\newcommand{\mechanical}{\raisebox{-0.2\height}{\includegraphics[width=0.018\textwidth]{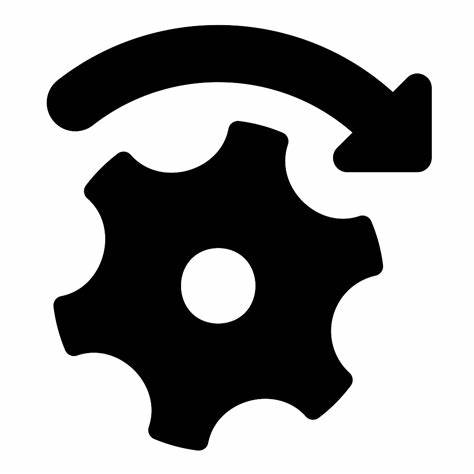}}}
\newcommand{\EM}{\raisebox{-0.2\height}{\includegraphics[width=0.016\textwidth]{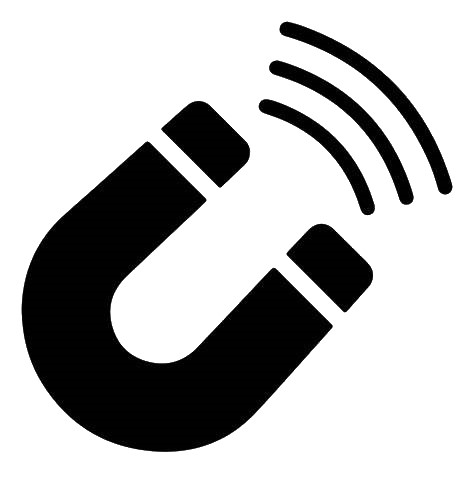}}}

\def\iconwidth{0.035\textwidth}
\newcommand{\CO}{\raisebox{-0.25\height}{\includegraphics[width=\iconwidth]{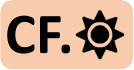}}}
\newcommand{\CA}{\raisebox{-0.25\height}{\includegraphics[width=\iconwidth]{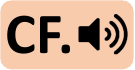}}}
\newcommand{\CE}{\raisebox{-0.25\height}{\includegraphics[width=\iconwidth]{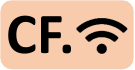}}}
\newcommand{\CM}{\raisebox{-0.25\height}{\includegraphics[width=\iconwidth]{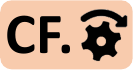}}}
\newcommand{\CT}{\raisebox{-0.25\height}{\includegraphics[width=\iconwidth]{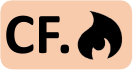}}}
\newcommand{\OA}{\raisebox{-0.25\height}{\includegraphics[width=\iconwidth]{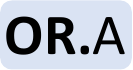}}}
\newcommand{\OF}{\raisebox{-0.25\height}{\includegraphics[width=\iconwidth]{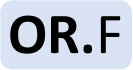}}}

\renewcommand{\c}{\Circle}
\newcommand{\C}{\CIRCLE}
\newcommand{\lc}{\LEFTcircle}
\def\cwidth{0.018\textwidth}
\setlength\dashlinedash{1pt}
\setlength\dashlinegap{2pt}

\begin{table*}[t!]
\caption{Taxonomy and mechanisms of sensor OOB vulnerability}
\centering
    \begin{tabular}{|c:c:c|c:c:c:c:c|c:c:c|c|c||c:c:c:c|}

    \hline
    \multicolumn{3}{|c|}{\textbf{Taxonomy}} & \multicolumn{5}{c|}{\textbf{Transducer}} & \multicolumn{3}{c|}{\textbf{Signal Cond.}} & \multirow{2}{*}{\makebox[\cwidth]{\textbf{Com.}}} & \multirow{2}{*}{\makebox[\cwidth]{\textbf{Pwr.}}} & \multicolumn{4}{c|}{\textbf{Unsecure Phase}} \\ \cline{1-11} \cline{14-17}
    Type & Subtype & Label & \makebox[\cwidth]{\faVolumeUp} & \makebox[\cwidth]{\faSun} & \makebox[\cwidth]{\faWifi} & \makebox[\cwidth]{\mechanical} & \makebox[\cwidth]{\faHotjar} & \makebox[\cwidth]{Amp.} & \makebox[\cwidth]{Fil.} & \makebox[\cwidth]{ADC} &  &  & \makebox[\cwidth]{\textbf{TS}} & \makebox[\cwidth]{\textbf{ECD}} & \makebox[\cwidth]{\textbf{P\&A}} & \makebox[\cwidth]{\textbf{T\&C}} \\ \hline
    \multirow{2}{*}{\textit{Out-of-range}} & Amplitude \textit{out-of-range} & \OA & \lc & \C & \lc & \lc & \lc & \C & \lc & \lc & \c & \c & \ding{52} & \ding{52} &  & \ding{52} \\ \cline{2-17}
    & Frequency \textit{out-of-range} & \OF & \C & \lc & \lc & \lc & \texttimes & \C & \C & \C & \texttimes & \C & \ding{52} & \ding{52} &  & \ding{52} \\ \hline
    \multirow{5}{*}{\textit{Cross-field}} & Acoustic signal \textit{cross-field} & \CA & \texttimes & \c & \c & \C & \c & \c & \c & \c & \c & \c & \ding{52} &  & \ding{52} & \ding{52} \\ \cline{2-17}
    & Optical signal \textit{cross-field} & \CO & \C & \texttimes & \C & \lc & \lc & \lc & \lc & \lc & \c & \c & \ding{52} &  & \ding{52} & \ding{52} \\ \cline{2-17}
    & EM signal \textit{cross-field} & \CE & \lc & \lc & \texttimes & \lc & \lc & \C & \C & \C & \C & \C & \ding{52} & \ding{52} & \ding{52} & \ding{52}\\ \cline{2-17}
    & Mechanical signal \textit{cross-field} & \CM & \C & \c & \c & \texttimes & \c & \c & \c & \c & \c & \c & \ding{52} &  & \ding{52} & \ding{52}\\ \cline{2-17}
    & Thermal signal \textit{cross-field} & \CT & \lc & \c & \c & \c & \texttimes & \lc & \lc & \lc & \c & \c & \ding{52} & \ding{52} & \ding{52} & \ding{52} \\ \hline
    
    \end{tabular}%
    \begin{tablenotes}
        \item \faVolumeUp\ Acoustic \quad \faSun\ Optical \quad \faWifi\ Electromagnetic (also includes electrical and magnetic) \quad \mechanical\ Mechanical \quad \faHotjar\ Thermal
        \item  \C\ Exploited \quad \lc\ Potential to be exploited  \quad \c\ Not exploitable \quad \texttimes\ Not applicable
        \item \textbf{TS} Transducer selection \quad \textbf{ECD} Electronic circuit design \quad \textbf{P\&A} Packaging and assembling \quad \textbf{T\&C} Testing and calibration
    \end{tablenotes}
    \label{tab:vuln_tax}
\end{table*}

\def\iconwidth{0.04\textwidth}
\subsubsection{Out-of-range Vulnerability}
We consider \textit{out-of-range} vulnerabilities from two aspects: amplitude and frequency. 

\begin{itemize}[leftmargin=15pt]
    \item \underline{\textit{Amplitude out-of-range}} (\OA). An amplitude \textit{out-of-range} signal impacts the output by the \textbf{saturation effect}. Transducers and signal conditioning circuits operate well within a predefined amplitude range. When the input signal exceeds this range, saturation occurs, resulting in signal clipping, i.e., $f'(x)=c\ (x\ge x_{max})$. For transducers, saturation is mainly due to the physical constraints of the transducer materials, e.g., the number of electron-hole pairs that can be generated in photovoltaic materials is limited~\cite{agostini1988photoelectric}. Thus, a high-intensity light can saturate an optical sensor~\cite{park2016ain}, leading to the maximum output. In signal conditioning circuits, saturation is generally caused by supply voltage limitations in active circuit elements. When the input is an alternating current (AC) signal, symmetric saturation introduces harmonic distortion, while asymmetric saturation also introduces a direct current (DC) bias. This mechanism has been exploited in amplifiers~\cite{trippel2017walnut}, and we posit it likely exists in filters and ADCs as well.
    
    \item \underline{\textit{Frequency out-of-range}} (\OF). A frequency \textit{out-of-range} signal can affect the output by nonlinearity, non-ideal cutoff, and aliasing effect. \textbf{Nonlinearity} is common in both transducers and signal processing circuits, and can be formulated as $f'(x)=a_0+a_1x+a_2x^2+\cdots$. In exploited cases, acoustic transducers and amplifiers have been shown to convert modulated high-frequency signals into low-frequency outputs via inter-modulation distortion (IMD)~\cite{zhang2017dolphinattack}. Similarly, nonlinear rectification in amplifiers can convert AC signals into DC offsets~\cite{tu2019trick}. \textbf{Non-ideal cutoff} refers to the imperfect frequency response of transducers and filters, which fails to effectively suppress high-intensity signals in the stop-band. As a result, we have $|F'(\omega)|>0\ (\omega\ge \omega_c)$, where $F'$ is the Fourier transform of $f'$ and $\omega_c$ is the cutoff frequency. Exploited cases include microphone transducers responding to ultrasound~\cite{zhang2017dolphinattack}, ultraviolet sensors reacting to visible lasers~\cite{shi2012comparative}, and low-pass filters failing to eliminate stop-band frequencies~\cite{wang2023volttack}. The \textbf{aliasing effect} occurs when ADCs receive input signals containing frequency components above the Nyquist frequency (i.e., $\omega>\omega_s/2$), where $\omega_s$ is the sampling rate. Then, the spectrum of $F'(\omega)$ contains aliasing frequencies $\omega_{a}=|\omega-k\omega_s|\ \textrm{where}\ k=\textrm{round}(\omega/\omega_s)$. As a result, high-frequency signals can be incorrectly interpreted as low-frequency components during sampling, effectively demodulating them into the output~\cite{kune2013ghost}. 
\end{itemize}

\subsubsection{Cross-field Vulnerability}
We consider \textit{cross-field} vulnerabilities in terms of signal modalities, i.e., acoustic, optical, electromagnetic, mechanical, and thermal. While optical, EM, and thermal signals all belong to the EM spectrum in the physical sense, we discuss them separately due to their fundamental differences in energy conversion mechanisms and propagation behaviors. Besides, since electric, magnetic, and EM signals are inherently coupled, we unify them under the term EM to simplify the taxonomy.
\begin{itemize}[leftmargin=15pt]
    \item \underline{\textit{Acoustic signal cross-field}} (\CA). Acoustic \textit{cross-field} signals can affect transducers by \textbf{resonance effect}. MEMS transducers used in motion sensors, such as accelerometers and gyroscopes, exhibit inherent resonant frequencies, making them sensitive to acoustic noise~\cite{dean2010characterization}. As a result, sound or ultrasound waves near the resonant frequency can induce high-intensity interference signals within the transducer~\cite{trippel2017walnut}, i.e., we have $g_2(z)=A\cos(\omega_at+\phi)$ where $\omega_a$ is the acoustic signal frequency.
    \item \underline{\textit{Optical signal cross-field}} (\CO). Optical \textit{cross-field} signals can affect transducers by \textbf{photoacoustic effect} and \textbf{photoelectric effect}. The photoacoustic effect converts light energy into mechanical vibrations, which can disturb sensors with diaphragm structures designed to detect such vibrations, such as MEMS microphones. These microphones have been shown to respond to amplitude-modulated light~\cite{sugawara2020light}, producing outputs like $g_2(z)=A[1+\cos(\omega_o+\phi)]$, where $\omega_o$ is the modulation frequency. The photoelectric effect, on the other hand, occurs when light liberates electrons from a material's surface, generating electric currents. Sensors with exposed conductive parts, such as MEMS barometers~\cite{tanaka2022laser}, are vulnerable to this effect, allowing the attacks to induce output bias under illumination, i.e., $g_2(z)=b$.

    \item \underline{\textit{Electromagnetic signal cross-field}} (\CE). Electromagnetic (EM) \textit{cross-field} signals can affect signal conditioning circuits by \textbf{antenna effect}, where conductive elements (especially wires) act as unintended antennas, receiving ($g_2(z)$) or transmitting ($g_1(x)$) electromagnetic radiation. This effect is especially pronounced when the conductor length approaches a quarter of the EM signal's wavelength, forming a resonant structure that efficiently couples EM energy into the circuit. This mechanism has been exploited by attackers in transducers~\cite{ni2023recovering}, amplifier input wires~\cite{tu2019trick}, ADCs~\cite{zhou2023dehirec}, and communication cables~\cite{jiang2023glitchhiker}.

    \item \underline{\textit{Mechanical signal cross-field}} (\CM). Just as acoustic signals can affect mechanical transducers, mechanical \textit{cross-field} signals can influence acoustic transducers also by \textbf{resonance effect}. A notable example is the injection of vibration signals directly into the diaphragm of an acoustic transducer~\cite{yan2020surfingattack}. However, mechanical signals must propagate through rigid media to reach the sensor, so their practicality for inducing OOB vulnerabilities is limited. Consequently, mechanical signals are not commonly exploited in this field.

    \item \underline{\textit{Thermal signal cross-field}} (\CT). Thermal \textit{cross-field} signals have not yet been directly exploited in attacks due to the difficulty of transmitting thermal energy over long distances and inducing swift temperature changes. Nevertheless, many sensor components are known to exhibit temperature sensitivity, such as \textbf{temperature drift} in amplifiers and other analog circuits~\cite{svoboda2013introduction}. Thus, we consider thermal \textit{cross-field} signals a potential but underexplored vector for inducing OOB vulnerabilities, which represents a notable gap in current research. 
\end{itemize}

Note that the above analysis is grounded in the energy conversion principles that have been exploited in existing attacks. In fact, other physical principles remain unexplored, which can be promising directions for future research.  For instance, the optical-pressure effect in mechanical transducers and thermal laser stimulation in thermal transducers may introduce new attack vectors, which are marked as potential in Table~\ref{tab:vuln_tax}. Our taxonomy of OOB vulnerability thus can serve as a reference for identifying such research gaps.

\subsection{Implications for Sensor Design and Vulnerability Testing}
Despite the inherent presence of OOB vulnerabilities in sensors, it is impractical to eliminate all non-idealities during the design phase due to the broad interdisciplinary nature of sensor engineering and the physical limitations of materials and components. Instead, it is essential to identify and prioritize non-idealities that may lead to exploitable vulnerabilities. To this end, our proposed vulnerability model and mechanism analysis can serve as a practical guide for designers to anticipate and mitigate security-critical issues throughout the sensor development lifecycle.

\subsubsection{Sensor Design}
Sensor design typically consists of three major stages: transducer selection, electronic circuit design, and packaging and assembling~\cite{webster2018measurement, tartagni2022electronic}. Non-ideal behaviors can arise at each of these stages and, while often treated as tolerable deviations in conventional performance metrics, they may be leveraged by attackers as vectors for OOB exploitation.
\textbf{a) Transducer Selection.} This phase is a critical source of potential OOB vulnerabilities. Designers often overlook unintended energy conversion pathways that arise from the inherent physical properties of selected materials or structural designs. For example, choosing MEMS-diaphragm-based transducers can expose even non-optical sensors to optical \textit{cross-field} signals due to the diaphragm's susceptibility to light-induced vibrations~\cite{sugawara2020light, tanaka2022laser}. These \textit{cross-field} responses are typically undocumented in datasheets, yet they can pose significant security risks.
\textbf{b) Electronic Circuit Design.} This phase introduces OOB vulnerabilities associated with signal conditioning circuits, power supplies, and communication interfaces. Non-idealities, such as saturation and nonlinearity, are inherent properties of electronic circuits. Designers typically define valid input ranges to mitigate their impact. However, in practice, it is infeasible to physically restrict the injection of OOB signals. As a result, sensor design alone is insufficient to fully eliminate OOB vulnerabilities in this phase, necessitating complementary system-level defense strategies, as discussed in Sec.~\ref{sec:defense}.
\textbf{c) Packaging and Assembly.} This phase can introduce additional \textit{cross-field} vulnerabilities due to insufficient shielding and structural alterations. Assembly may modify the mechanical structure and electrical characteristics of the sensor, potentially introducing new mechanical resonance or electromagnetic coupling frequencies. For instance, through consulting with system developers, we found that MEMS vibration sensors mounted at certain locations on a motherboard can be affected by system-level resonance frequencies, leading to inaccurate measurements.

\subsubsection{Vulnerability Testing}
Existing sensor testing is insufficient since it primarily focuses on the sensor's accuracy. Key characteristics such as functionality, sensitivity, and linearity are typically evaluated using \textit{in-band} signals, while the potential effects of OOB vulnerabilities are largely overlooked. Although some tests, such as electromagnetic compatibility (EMC) testing, consider EM \textit{cross-field} signals, they are limited to specific frequency ranges and cannot effectively cover the broader spectrum of potential attack signals~\cite{wang2023volttack}. Consequently, such tests are inadequate for detecting OOB vulnerabilities, which also remain undocumented in sensor datasheets. That said, not all OOB signals warrant mitigation. Whether an OOB signal should be considered a threat depends on whether the sensor’s transduction principle allows it to respond to that signal, and whether the transduction process is efficient enough to be exploited. For example, although MEMS sensors contain conductive components in both the transducer and internal circuitry, typical EMI signals rarely succeed in attacking them. We suppose this is due to the highly integrated, micrometer-scale internal structure of MEMS devices, which would require electromagnetic waves with micrometer-scale wavelengths, i.e., in the terahertz (THz) or near-infrared range, to interact effectively. Since such frequencies border on or fall within the optical spectrum, conventional EMC testing for these bands is neither practical nor necessary for MEMS transducers and their associated conditioning circuits.

In conclusion, we argue that \textit{security-aware sensor design} must evolve beyond traditional performance-centric paradigms. By incorporating OOB vulnerability taxonomy early in the design process, engineers can better anticipate potential attack vectors and adopt mitigation strategies such as shielding, filtering, redundancy, or cross-domain isolation, which will be further discussed in Sec.~\ref{sec:defense}. Rather than attempting to eliminate all nonidealities, designers should focus on identifying and mitigating those most likely to be adversarially exploitable. Our formalization provides the foundation for this security-aware approach to sensor security.

\setlength\dashlinedash{1pt}
\setlength\dashlinegap{2pt}

\newcommand{\one}{\raisebox{-0.2\height}{\includegraphics[width=0.015\textwidth]{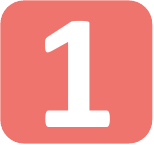}}}
\newcommand{\two}{\raisebox{-0.2\height}{\includegraphics[width=0.015\textwidth]{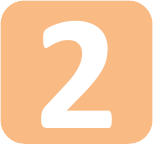}}}
\newcommand{\three}{\raisebox{-0.2\height}{\includegraphics[width=0.015\textwidth]{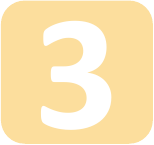}}}
\newcommand{\four}{\raisebox{-0.2\height}{\includegraphics[width=0.015\textwidth]{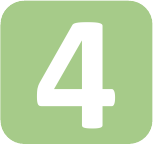}}}
\newcommand{\five}{\raisebox{-0.2\height}{\includegraphics[width=0.015\textwidth]{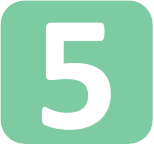}}}
\newcommand{\NA}{\raisebox{-0.2\height}{\includegraphics[width=0.015\textwidth]{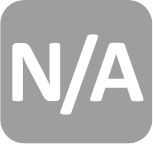}}}
\newcommand{\force}{\raisebox{-0.15\height}{\includegraphics[width=0.015\textwidth]{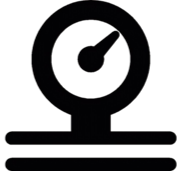}}}
\newcommand{\lidar}{\raisebox{-0.15\height}{\includegraphics[width=0.016\textwidth]{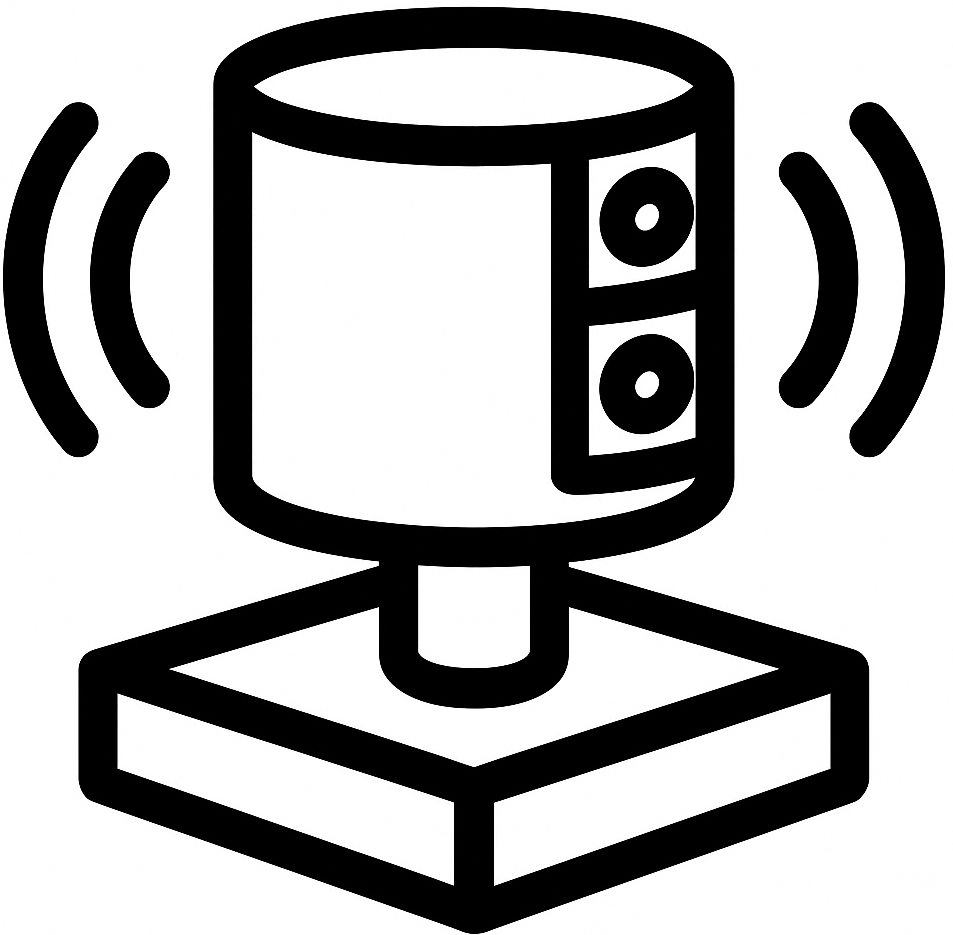}}}
\newcommand{\Infrared}{\raisebox{-0.15\height}{\includegraphics[width=0.016\textwidth]{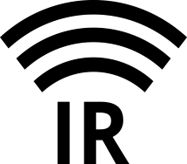}}}
\newcommand{\touch}{\raisebox{-0.15\height}{\includegraphics[width=0.016\textwidth]{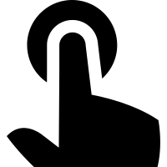}}}
\newcommand{\VA}{\raisebox{-0.15\height}{\includegraphics[width=0.016\textwidth]{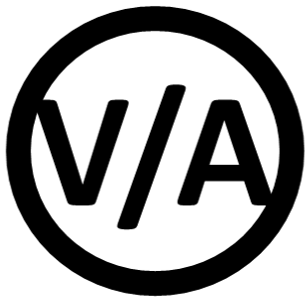}}}
\newcommand{\humidity}{\raisebox{-0.15\height}{\includegraphics[width=0.016\textwidth]{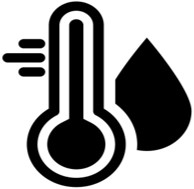}}}
\newcommand{\acc}{\raisebox{-0.25\height}{\includegraphics[width=0.02\textwidth]{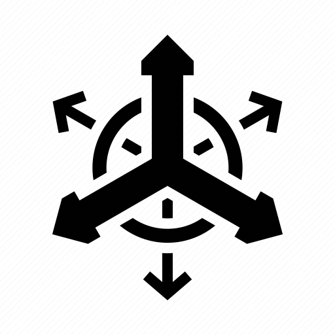}}}

\def\iconwidth{0.035\textwidth}

\begin{table*}[t!]
\caption{Systematization of sensor-level attacks}
\resizebox{\textwidth}{!}{%
\begin{tabular}{|c|c|c:c:c|c:c:c|c:c:c:c|l|}
\hline
\multirow{2}{*}{\textbf{\makecell{\textbf{Attack} \\ \textbf{Signal}}}} & \multirow{2}{*}{\textbf{\makecell{\textbf{Target} \\ \textbf{Sensor}}}} & \multicolumn{3}{c|}{\textbf{Attack Goal}} & \multicolumn{3}{c|}{\textbf{Attack Path}} & \multicolumn{4}{c|}{\textbf{Attack Practicality}} & \multicolumn{1}{c|}{\multirow{2}{*}{\textbf{Paper}}} \\ \cline{3-12}
 &  & \makebox[0.02\textwidth]{DoS} & \makebox[0.02\textwidth]{Spoof} & \makebox[0.02\textwidth]{Snoop} & Parameter & Transduction & Process & Knowl. & Range & Cost & Size &  \\ \hline
\multirow{4}{*}{\rotatebox{90}{\makecell{\textbf{Sound}/\\\textbf{Ultrasound}}}} & \faMicrophone* & \ding{51} & \ding{51} &  & AM/PWM/FM & Trans.\OF & Amp.\OF & \faSquare & \one$\sim$\five & \two$\sim$\five & \one$\sim$\five & \cite{roy2017backdoor, shen2019jamsys, ji2021capspeaker, jiang2023capspeaker, roy2018inaudible, yan2019feasibility, zhang2017dolphinattack, yan2020surfingattack, li2023echoattack, ning2025portable, li2023enrollment, li2024inaudible, zheng2023silent} \\ \cdashline{2-13}
 & \multirow{2}{*}{\acc} & \ding{51} & \ding{51} &  & Sine & Trans.\CA & Amp.\OA/Fil.\OF/ADC\OF & \faSquare[regular] & \four & \three$\sim$\five & \one$\sim$\three & \cite{trippel2017walnut, tu2018injected, son2015rocking, ji2021poltergeist, zhu2023tpatch} \\ \cdashline{3-13}
 &  &  &  & \ding{51} & / & Trans.\CA & / & \faSquare & \two & \five & \five & \cite{anand2018speechless, han2017pitchln, ba2020learning, hu2022accear, matovu2019kinetic, shi2021face, zhang2015accelword, michalevsky2014gyrophone, kwong2019hard} \\ \cdashline{2-13}
 & \force &  & \ding{51} &  & Sine & Trans.\CA & / & \faSquare[] & \one & \four & \five & \cite{barua2022wolf} \\ \hline
\multirow{3}{*}{\rotatebox{90}{\textbf{Laser}}} & \faMicrophone* &  & \ding{51} &  & AM/PWM & Trans.\CO & / & \faSquare & \three$\sim$\five & \three & \five & \cite{cyr2021lasers, sugawara2020light, shi2024laser, zhang2024laseradv} \\ \cdashline{2-13}
 & \force &  & \ding{51} &  & AM & Trans.\CO & / & \faSquare & \NA & \one & \one &  \cite{tanaka2022laser} \\ \cdashline{2-13}
 & \faCamera\ \lidar\ \Infrared & \ding{51} &  &  & Const. & Trans.\OA & / & \faSquare & \three$\sim$\five & \two$\sim$\five & \one$\sim$\five & \cite{petit2015remote, shin2017illusion, park2016ain} \\ \hline
\multirow{10}{*}{\rotatebox{90}{\makecell{\textbf{Radiated}\\\textbf{EMI}}}} & \faMicrophone* & \ding{51} & \ding{51} &  & AM & Pre-Amp.\CE & Amp.\OF & \faSquare & \one$\sim$\three & \four & \one$\sim$\five & \cite{dai2023inducing, kasmi2015iemi, kune2013ghost} \\ \cdashline{2-13}
 & \multirow{2}{*}{\faCamera} &  & \ding{51} &  & AM/PM & Pre-ADC\CE & / & \faSquare[regular] & \two & \one$\sim$\two & \one$\sim$\three & \cite{kohler2022signal, Ren2025GhostShotMT} \\ \cdashline{3-13}
 &  &  &  & \ding{51} & / & Comm.\CE & / & \faSquare & \two & \three & \one & \cite{long2024emeye} \\ \cdashline{2-13}
 & \lidar &  & \ding{51} &  & AM & Pre-Amp.\CE & Amp.\OF/ADC\OF & \faSquare[regular] & \four & \one & \one & \cite{jin2024phantomlidar} \\ \cdashline{2-13}
 & \touch & \ding{51} & \ding{51} & \ding{51} & Sine & Trans.\CE & / & \faSquare/\faSquare[regular] & \one$\sim$\two & \one$\sim$\five & \one$\sim$\five & \cite{maruyama2017poster, maruyama2019tap, shan2022invisible, wang2022ghosttouch, wang2024analyzing, jin2021periscope} \\ \cdashline{2-13}
 & \faFingerprint &  &  & \ding{51} & / & Trans.\CE & / & \faSquare & \one & \four & \five & \cite{ni2023recovering} \\ \cdashline{2-13}
 & \multirow{2}{*}{\VA} &  & \ding{51} &  & Sine & Pre-ADC\CE & ADC\OF & \faSquare[regular] & \one & \three & \three & \cite{dayanikli2020electromagnetic} \\ \cdashline{3-13}
 &  &  & \ding{51} &  & AM & Pre-Amp.\CE & Amp.\OF & \faSquare[regular] & \three & \one & \one & \cite{yang2024rethink} \\ \cdashline{2-13}
 & \faThermometerHalf &  & \ding{51} &  & Sine & Pre-Amp.\CE & Amp.\OF & \faSquare[regular] & \five & \one & \one & \cite{tu2019trick} \\ \cdashline{2-13}
 & \Infrared &  &  & \ding{51} & / & Comm.\CE & / & \faSquare & \one & \one & \three & \cite{EMIRIS} \\ \hline
\multirow{6}{*}{\rotatebox{90}{\makecell{\textbf{Conducted}\\\textbf{EMI}}}} & \faThermometerHalf\ \humidity\ \acc\ \force & \ding{51} & \ding{51} &  & Sine & Pwr.\OF & Amp.\OF/ADC\OF & \faSquare[regular] & \NA & \one$\sim$\two & \five & \cite{wang2023volttack} \\ \cdashline{2-13}
 & \multirow{2}{*}{\faMicrophone*} &  & \ding{51} &  & AM & Pwr.\OF & Amp.\OF & \faSquare[regular] & \five & \three & \three & \cite{esteves2018remote} \\ \cdashline{3-13}
 &  &  & \ding{51} &  & Const./Sine/AM & Pwr.\OF & Amp.\OF/ADC\OF & \faSquare[regular] & \five & \two & \one & \cite{jiang2024powerradio} \\ \cdashline{2-13}
 & \faCamera &  & \ding{51} &  & Sine & Pwr.\OF & Amp.\OF & \faSquare[] & \NA & \four & \three & \cite{jiang2025vphanton} \\ \cdashline{2-13}
 & \multirow{2}{*}{\touch} & \ding{51} & \ding{51} &  & Sine/AM & Pwr.\OF & Trans.\CE & \faSquare[regular] & \three & \three$\sim$\four & \one$\sim$\three & \cite{jiang2022wight, jiang2023marionette, zhu2022powertouch} \\ \cdashline{3-13}
 &  &  &  & \ding{51} & / & Pwr.\CE & / & \faSquare & \NA & \five & \five & \cite{cronin2021charger}\\
 \hline
\end{tabular}%
}
\begin{tablenotes}
    \item \faMicrophone*\ Microphone \quad \acc\ Motion sensor \quad \force\ Force/pressure sensor \quad \faCamera\ Image sensor \quad \lidar\ Lidar \quad \Infrared\ Infrared sensor \quad \touch\ Touchscreen \quad\faFingerprint\ Fingerprint sensor \quad \VA\ Voltage/current sensor \quad \faThermometerHalf\ Thermometer \quad \humidity\ Humidity sensor \quad \OA \OF \CA \CO \CE\ Exploited OOB vulnerability mechanisms
    \item \faSquare\ Black-box \quad \faSquare[]\ White-box \quad \one\two\three\four\five\ Higher level indicates higher practicality \quad \NA\ Not available
\end{tablenotes}
\label{tab:attacks}
\end{table*} 
\def\iconwidth{0.04\textwidth}

\section{Sensor-level Attack Systematization} \label{sec:sensor-level}
In this section, we systematically analyze how OOB sensor vulnerabilities enable \textit{sensor-level} attacks. While numerous attacks exist, they remain fragmented as isolated cases with disparate attack scenarios and threat models of varying severity, which hinders the generalization of the attack methods across different sensors. Moreover, attackers tend to adopt favorable threat models for attack success that may exaggerate the real-world practicality of attacks~\cite{walker2021sok}. Thus, in this section, we conduct a comprehensive analysis of existing literature, identify common mechanisms among different studies, and evaluate the practicality of each attack. 

\subsection{Systematization Methodology}
We systematize sensor-level attacks in terms of \textit{attack signal}, \textit{target sensor}, \textit{attack path}, \textit{attack goal}, and \textit{attack practicality}, as shown in Table~\ref{tab:attacks}.

\subsubsection{Attack Signal} We classify the attack signals into four modalities. \textbf{a) Sound/Ultrasound}, \textbf{b) Laser}, \textbf{c) Radiated EMI}, and \textbf{d) Conducted EMI}. Note that in our context, the term \textit{attack signal} refers to both  malicious signals actively emitted by attackers and passive side-channel leakage from sensors. This helps to unify the commonality of various attack vectors that exploit OOB vulnerabilities.

\subsubsection{Attack Goal} We identify three attack goals in existing work. 
\textbf{a) Denial-of-Service (DoS):} The attacker aims to make the measurement unavailable by overwhelming it with powerful noise. 
\textbf{b) Spoofing:} The attacker spoofs the sensors to produce seemingly legitimate but erroneous measurements. 
\textbf{c) Snooping:} The attacker exploits the side-channel leakage of a sensor to recover private information.

\subsubsection{Attack Path} The attack path includes three stages. \textbf{a) Signal parameter:} describes how attack signals are modulated. This includes constant signals (Const.), single-frequency wave (Sine), amplitude modulation (AM), frequency modulation (FM), phase modulation (PM), and pulse width modulation (PWM). \textbf{b) Signal transduction:} describes how attack signals are either injected into or emitted from the target sensor.\textbf{ \textbf{c)} Signal processing:} describes how these signals are manipulated by the sensor's internal circuitry. For each stage, we identify the underlying OOB vulnerability mechanisms (\OA\OF\CA\CO\CE) and the associated sensor components (Sec.~\ref{sec:mechanisms}). The \textbf{Pre-} prefix denotes that the signal is injected into the front-end wires of the components.

\renewcommand{\one}{\raisebox{-0.2\height}{\includegraphics[width=0.02\textwidth]{icon/num1.pdf}}}
\renewcommand{\two}{\raisebox{-0.2\height}{\includegraphics[width=0.02\textwidth]{icon/num2.pdf}}}
\renewcommand{\three}{\raisebox{-0.2\height}{\includegraphics[width=0.02\textwidth]{icon/num3.pdf}}}
\renewcommand{\four}{\raisebox{-0.2\height}{\includegraphics[width=0.02\textwidth]{icon/num4.pdf}}}
\renewcommand{\five}{\raisebox{-0.2\height}{\includegraphics[width=0.02\textwidth]{icon/num5.pdf}}}
\renewcommand{\force}{\raisebox{-0.15\height}{\includegraphics[width=0.02\textwidth]{icon/force.png}}}
\renewcommand{\lidar}{\raisebox{-0.15\height}{\includegraphics[width=0.02\textwidth]{icon/Lidar2.png}}}
\renewcommand{\Infrared}{\raisebox{-0.15\height}{\includegraphics[width=0.02\textwidth]{icon/IR.png}}}
\renewcommand{\touch}{\raisebox{-0.15\height}{\includegraphics[width=0.02\textwidth]{icon/touchscreen.png}}}
\renewcommand{\VA}{\raisebox{-0.15\height}{\includegraphics[width=0.02\textwidth]{icon/VA.png}}}
\renewcommand{\humidity}{\raisebox{-0.15\height}{\includegraphics[width=0.02\textwidth]{icon/humidity.jpg}}}
\renewcommand{\acc}{\raisebox{-0.24\height}{\includegraphics[width=0.025\textwidth]{icon/accelerometer.png}}}

\subsubsection{Attack Practicality} We evaluate attack practicality through four dimensions.
\textbf{a) Prior knowledge:} black-box (\faSquare) indicates the attack requires no sensor-specific knowledge, while white-box (\faSquare[]) indicates the need for detailed sensor parameters such as sensor model, sensitive frequencies, or hardware characteristics. Note that the knowledge here is sensor-level rather than system-level. 
\textbf{b) Attack range:} we classify the maximum demonstrated attack range into five intervals: \one: $\leq$0.1m, \two: (0.1, 1m], \three: (1, 5m], \four: (5, 10m], and \five:$>$10m. 
\textbf{c) Attack device cost:} the price to set up the attack device is divided into \one: $>$10,000\$, \two: (5,000\$, 10,000\$], \three: (1,000\$, 5,000\$], \four: (100\$, 1,000\$], and \five: $<$100\$. 
\textbf{d) Attack device size:} indicates the ease and stealth to perform an attack~\cite{xu2023sok}, and we classify it into \one: fixed installation, \three: backpack-portable, and \five: hand-held device based on their size. In summary, a higher score indicates a higher practicality, which corresponds to a higher attack threat level.

\subsection{Review of Existing Work}

\subsubsection{Attacks by Sound/Ultrasound}

\textbf{a) Acoustic sensors (\faMicrophone*).} Attackers can conduct DoS and spoofing attack by exploiting the sensor's response to ultrasonic signals (\OF) through the injection of frequency \textit{out-of-range} ultrasounds~\cite{roy2017backdoor, shen2019jamsys, ji2021capspeaker, jiang2023capspeaker, roy2018inaudible, yan2019feasibility, zhang2017dolphinattack, yan2020surfingattack, li2023echoattack, ning2025portable, li2023enrollment, li2024inaudible, zheng2023silent}. By applying amplitude modulation, ultrasounds can be demodulated into \textit{in-band} audible commands or noises via the nonlinear IMD (\OF) of the amplifier. Such attacks are considered black-box (\faSquare) since nonlinearities are common in microphones and cannot be eliminated. The longest attack range is around 20m (\five), as demonstrated in~\cite{yan2019feasibility}. Most of these attacks have low device costs (\four~\cite{roy2017backdoor, shen2019jamsys, yan2019feasibility, zhang2017dolphinattack}, \five~\cite{ji2021capspeaker, jiang2023capspeaker}) and good portability (\three~\cite{roy2018inaudible, yan2020surfingattack}, \five~\cite{roy2017backdoor, shen2019jamsys, ji2021capspeaker, jiang2023capspeaker, zhang2017dolphinattack, ning2025portable}). However, there remains a trade-off between attack range and device portability.
\textbf{b) Non-acoustic sensors (\acc\ \force).} Attackers can launch DoS and spoofing attacks via \textit{cross-field} injections of sound or ultrasound, which induce mechanical resonance in the sensor structure (\CA)~\cite{trippel2017walnut, tu2018injected, son2015rocking, ji2021poltergeist, zhu2023tpatch, barua2022wolf}. These induced signals can be further exploited through mechanisms such as amplifier saturation (\OA), imperfect filter cutoff (\OF), and ADC aliasing (\OF). Since effective attacks require knowledge of the target sensor’s resonant frequency, they are categorized as white-box attacks (\faSquare[]). The maximum demonstrated attack range is 7.7m (\four~\cite{tu2018injected}). However, such attacks typically require an audio amplifier to improve the attack signal strength, limiting their portability (\one~\cite{trippel2017walnut, tu2018injected}). In addition, attackers can perform snooping attacks by using motion sensors to capture subtle mechanical vibrations induced by sound waves~\cite{anand2018speechless, han2017pitchln, ba2020learning, hu2022accear, matovu2019kinetic, shi2021face, zhang2015accelword, michalevsky2014gyrophone, kwong2019hard}, enabling the reconstruction of private speech information. Although such attacks require no extra devices, they suffer from limited attack range (\three~\cite{anand2018speechless}, \two~\cite{han2017pitchln, zhang2015accelword, kwong2019hard}) due to the low power of human voice signals.

\subsubsection{Attacks by Laser}
\textbf{a) Optical sensors (\faCamera\ \lidar\ \Infrared).}
Attackers can launch DoS attacks by exploiting the saturation effect (\OA) of optical transducers through the injection of amplitude \textit{out-of-range} laser signals~\cite{petit2015remote, shin2017illusion, park2016ain}. These attacks are classified as black-box (\faSquare), as saturation is a common and predictable characteristic of optical sensors, requiring no sensor-level knowledge. Thanks to the strong directivity and low divergence of laser beams, signal attenuation over distance is minimal, enabling long-range attacks exceeding 10 meters (\five~\cite{shin2017illusion, park2016ain}) to be carried out using low-cost (\five~\cite{petit2015remote, park2016ain}) and highly portable laser devices (\five~\cite{petit2015remote, park2016ain}). However, in real-world scenarios, maintaining a stable laser focus on the target sensor over a long distance often requires fixed setups, which can reduce the practical portability of the attack. Note that here we do not consider lidar spoofing attacks~\cite{sun2020towards, cao2021invisible, hallyburton2022security, jin2023pla, cao2023you,   sato2024lidar, sato2025realism} since the signals are \textit{in-band}.
\textbf{b) Non-optical sensors (\faMicrophone*\ \force).}
Spoofing attacks have also been demonstrated against non-optical sensors such as MEMS microphones~\cite{sugawara2020light, cyr2021lasers, shi2024laser, zhang2024laseradv} and pressure sensors~\cite{tanaka2022laser}, by leveraging the photoacoustic (\CO) and photoelectric (\CO) effects, respectively. For microphones, attackers amplitude-modulate acoustic signals onto laser beams to induce signal injection, achieving ranges of up to 25 meters (\five~\cite{sugawara2020light}). The necessary attack devices, including laser drivers, signal repeaters, etc, can be obtained at moderate cost (\three~\cite{sugawara2020light, shi2024laser, zhang2024laseradv}). However, similar to optical DoS attacks, long-range attacks typically require stable and fixed installations, such as tripods~\cite{sugawara2020light}, limiting portability.

\subsubsection{Attacks by Radiated EMI}
Attackers can utilize signal generators, software radios, power amplifiers, and antennas to emit malicious signals that couple into vulnerable components, such as transducers~\cite{maruyama2017poster, maruyama2019tap, shan2022invisible, wang2022ghosttouch, wang2024analyzing}, amplifier front-end wiring~\cite{dai2023inducing, kasmi2015iemi, kune2013ghost, jin2024phantomlidar, yang2024rethink, tu2019trick}, and ADC front-end wiring~\cite{kohler2022signal, Ren2025GhostShotMT, dayanikli2020electromagnetic}, leading to DoS or spoofing behaviors. The injected signals can be further manipulated by the nonlinearity of amplifiers (\OF)~\cite{dai2023inducing, kasmi2015iemi, kune2013ghost, jin2024phantomlidar, yang2024rethink, tu2019trick} or the aliasing effect of ADCs (\OF)~\cite{jin2024phantomlidar, dayanikli2020electromagnetic}. In addition to active injection, attackers can also employ antennas and spectrum analyzers to passively receive side-channel EM emissions from transducers~\cite{jin2021periscope, ni2023recovering} and communication wires~\cite{long2024emeye, EMIRIS} to launch snooping attacks. These attacks share a common mechanism, i.e., all conductors in sensors exhibit the antenna effect (\CE). However, effective EMI injection requires knowledge of the coupling frequencies of the target conductors, which are typically obtained by preliminary frequency sweeping tests. Thus, EMI injection attacks are classified as white-box (\faSquare[]). 
Despite their broad applicability, the practical implementation of these attacks is limited. On the one hand, the attack range is generally short (\one~\cite{dai2023inducing, maruyama2017poster, maruyama2019tap, shan2022invisible, wang2022ghosttouch, wang2024analyzing, ni2023recovering, dayanikli2020electromagnetic, EMIRIS}, \two~\cite{jin2021periscope}) due to the need for precise alignment with vulnerable components or the inherently weak power of emitted side-channel signals. On the other hand, the devices required, including high-end RF gear and precision antennas, are generally expensive (\one~\cite{Ren2025GhostShotMT, wang2024analyzing, yang2024rethink, tu2019trick, EMIRIS}, \two~\cite{kohler2022signal, wang2022ghosttouch}) and bulky (\one~\cite{kasmi2015iemi, kune2013ghost, long2024emeye, Ren2025GhostShotMT, jin2024phantomlidar, maruyama2017poster, maruyama2019tap, shan2022invisible, wang2022ghosttouch, wang2024analyzing, yang2024rethink, tu2019trick}), reducing the portability and stealth of such attacks.

\subsubsection{Attacks by Conducted EMI}
Conducted EMI propagates along transmission lines such as power or ground cables, enabling attackers to induce false measurements in connected sensors (\faThermometerHalf\ \humidity\ \acc\ \force\ \faMicrophone*\ \touch\ \faCamera). The primary mechanism that facilitates the injection is the non-ideal filtering (\OF) of the power supply noise. Furthermore, these attacks can exploit other mechanisms, such as the nonlinearity of amplifiers (\OF)~\cite{wang2023volttack,esteves2018remote,jiang2024powerradio}, the non-ideal cutoff of filters (\OF)~\cite{wang2023volttack}, and aliasing effect of ADCs (\OF)~\cite{wang2023volttack,esteves2018remote,jiang2024powerradio}. Because the attack signal can propagate stably along cables, these attacks are effective in relatively long ranges (\five~\cite{esteves2018remote,jiang2024powerradio}, \three~\cite{jiang2022wight,jiang2023marionette,zhu2022powertouch}). One representative approach involves manipulating sensor outputs by injecting fluctuations into the power supply~\cite{wang2023volttack,jiang2025vphanton,esteves2018remote}, exploiting the sensitivity of internal sensor components to voltage variations. Another technique targets the ground line, where attackers inject malicious signals by exploiting circuit asymmetries~\cite{jiang2022wight,jiang2023marionette,zhu2022powertouch,jiang2024powerradio}. However, similar to radiated EMI attacks, conducted EMI attacks typically require prior knowledge of the vulnerable coupling frequency. As a result, they are classified as white-box attacks (\faSquare[]).

\subsection{Research Gaps and Future Directions}

\subsubsection{Enhancing the attack practicality}
Attack practicality is critical to assessing real-world threats, but often overlooked. Our analysis above reveals that acoustic/ultrasonic and radiated EMI attacks typically suffer from limited attack ranges and bulky, expensive attack devices. Thus, we highlight two directions for improving practicality as follows.

\begin{itemize}[leftmargin=10pt]
    \item \underline{\textit{Extending attack range.}} While increasing signal power may improve range, it always introduces trade-offs: higher device cost, larger device size, and safety risks to attackers. Moreover, for acoustic signals, higher power may cause audible leakage due to nonlinearities in speakers and air propagation~\cite{naugolnykh1998nonlinear, ning2025portable} and compromise attack stealthiness. Instead, attackers can improve signal directionality to boost received power at the target. Promising techniques include using phased arrays, acoustic metameterials~\cite{ning2025portable} and metasurfaces~\cite{assouar2018acoustic}, and metasurface antennas~\cite{badawe2016true} to enhance the signal directionality.
    \item \underline{\textit{Optimizing attack devices.}} Many attacks still rely on laboratory-grade signal generators and power amplifiers, despite their functionalities often exceeding the actual needs of signal injection. For example, attacks that use single-frequency signals (denoted as Sine in Table~\ref{tab:attacks}) do not require full-featured signal generators. In such cases, smartphones can serve as substitutes for generating acoustic signals, while phone-sized USRPs can be used for EM signal generation. Similarly, power amplifiers can be tailored to the signal’s frequency band. For example, attackers can employ dedicated audio amplifiers and RF-specific amplifiers instead of bulky and general-purpose amplifiers.
\end{itemize}

\subsubsection{New Attack Surfaces}
Based on the analysis in Table~\ref{tab:vuln_tax} and ~\ref{tab:attacks}, we find some potential attack surfaces that have not yet been explored, which are summarized below.
\begin{itemize}[leftmargin=10pt]
    \item \underline{\textit{Acoustic-based attacks on MEMS components.}}  
    Acoustic attacks have largely targeted MEMS-based transducers (e.g., accelerometers), due to their ultrasonic resonance. Other MEMS sensors (e.g., magnetometers, thermometers) and MEMS-based components such as clock oscillators~\cite{liu2024timetravel} may also be susceptible to acoustic injection, especially in smart sensors with integrated timing circuits. Thus, it's also possible to determine the presence of resonance frequencies by sweeping test and to further investigate their effect on sensor measurements.
    \item \underline{\textit{Laser-based attacks on signal conditioning circuits.}}  
    Laser-based attacks mainly target transducers, overlooking signal conditioning components. Yet, prior studies show that lasers can disrupt digital ICs~\cite{skorobogatov2003optical, nagata2021physical} via photoelectric effects. Given that amplifiers, filters, and ADCs also use transistors, these circuits may similarly be vulnerable. Researchers can inject lasers into circuits and analyze their effect on sensor measurements.
    \item \underline{\textit{EM-based snooping attacks on signal conditioning circuits.}}  
    Existing EM-based snooping attacks focus on transducers and communication wires. However, studies have also shown that EM side-channel emissions from ADCs~\cite{zhou2023dehirec} and clock circuitry~\cite{ramesh2022ticktock} can also leak sensor activity. These findings suggest that signal conditioning circuits could expose additional side-channel attack surfaces. Future work could investigate how to capture these emissions using antennas and reconstruct sensitive measurement data using AI-based inference techniques, such as generative adversarial networks (GANs)~\cite{hu2022accear}.
    \item \underline{\textit{Attacks on membrane-structured sensors.}}  
    Sensors with exposed membrane structures are always vulnerable, such as microphones~\cite{zhang2017dolphinattack} and barometers~\cite{tanaka2022laser}.
    Flexible sensors used in wearables and biomedical applications such as photoacoustic spectroscopy IR detectors and thermal-film sensors also feature exposed membranes. Due to lightweight design constraints, these sensors often lack proper shielding, making them likely susceptible to optical and EM interference. Thus, researchers can also study the effect of lasers and EM signals on these sensors.
    \item \underline{\textit{Self-attack paradigm.}}  
    For scenarios where attack signals are hard to reach and focus, such as high speed drones, we envision a new attack paradigm called \textit{self-attack}, where attackers can exploit a device’s own components to compromise its sensors. This idea is inspired by previous studies showing that onboard speakers can emit malicious audio targeting local microphones~\cite{xia2023near}, and capacitors can generate ultrasonic signals via the inverse piezoelectric effect~\cite{ji2021capspeaker}. Therefore, it is an interesting research direction to study, for example, how to make motors generate acoustic signals and thus interfere with motion sensors.
\end{itemize}

\begin{figure*}
    \centering
    \includegraphics[width=\linewidth]{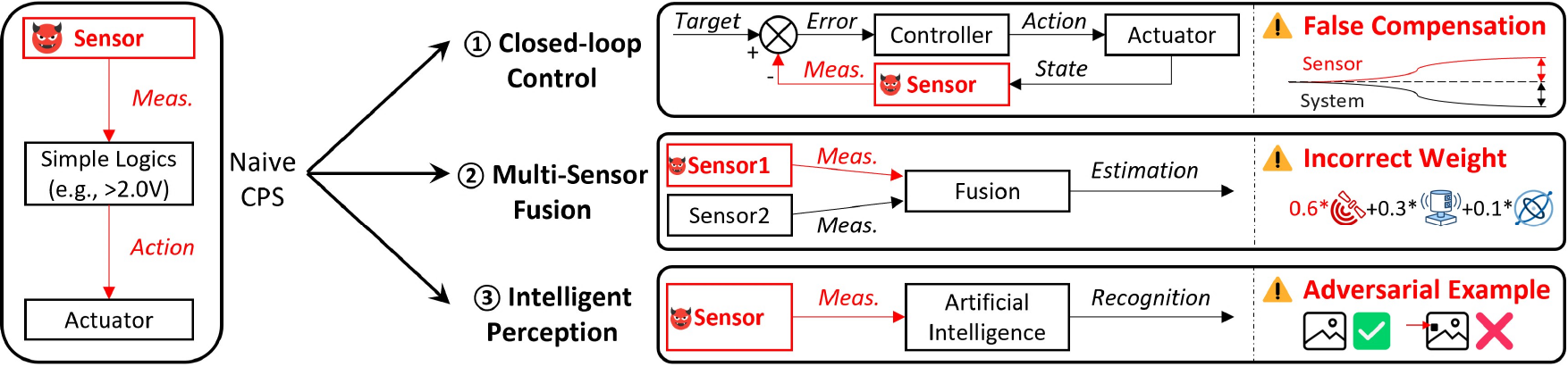}
    \caption{Implications of sensor failures on naive CPS and three typical advanced CPS. The red color indicates the compromised elements. While naive CPS behavior is directly affected, advanced CPS remain resilient unless a specific vulnerability is exploited.}
    \label{fig:system}
\end{figure*}

\section{System-level Implications} \label{sec:sys}

In this section, we expand the implications of attacks from individual sensors to CPS. While sensors are critical components of CPS, the intuition that \emph{sensor-level failures lead to system-level consequences} may not always be true. To understand \emph{why} sensor-level failures \emph{can/cannot} lead to system-level consequences, we identify three key characteristics of modern CPS: \textbf{a) closed-loop control}, \textbf{b) multi-sensor fusion}, and \textbf{c) intelligent perception}, as shown in Fig.~\ref{fig:system}. These characteristics complicate the relationship between sensor failures and system behavior. In contrast, a naive CPS, defined by open-loop control, single-sensor input, and no intelligent processing, reacts directly to sensor data, such as a light switch based on ambient brightness.
Although CPSs are diverse and encompass many other characteristics, we focus on these three because they are the most prevalent and influential when it comes to the system-level impact of sensor attacks. More importantly, this section does not delve into specific attack methods, as discussed in Sec.~\ref{sec:sensor-level}. Instead, we take sensor failures as a basic assumption for analyzing their implications at the system level.

\subsection{Implications on Closed-loop Control}

For CPS requiring accurate control and robustness to external disturbances or internal variations, it is pervasive to adopt a closed-loop (CL) architecture, as shown in Fig.~\ref{fig:system}, where the sensor acts as the feedback for continuous control adjustments~\cite{nise2020control}. Compared to open-loop systems, the sensor feedback mechanism in CL systems complicates the relationship between sensor failures and system consequences.

\subsubsection{System's Strength} Since CL systems are less sensitive to noise and disturbances in the environment than open-loop systems~\cite{nise2020control}, they can, to some extent, resist noisy sensor readings that are equivalent to environmental disturbances. For example, CL flight control systems are immune to most sinusoidal perturbations (e.g., $>$5 Hz~\cite{jeong2023rocking}) induced by resonant acoustic signal injection on MEMS gyroscopes~\cite{son2015rocking}.

Moreover, CL systems use sensors to measure states that are largely affected by previous actions, making the sensor measurements somewhat predictable. This characteristic benefits anomaly detection mechanisms, e.g., monitoring the discrepancy between the expected state and sensor feedback. A substantial discrepancy may trigger alarming~\cite{tu2019trick} or fail-safe logic~\cite{kim2024systematic}. A line of research develops more advanced anomaly detection methods. \citet{choi2018detecting} proposed to detect malicious sensor readings by monitoring control invariants obtained from a linear model of the victim system. \citet{quinonez2020savior} further improved this method by introducing non-linear control invariants and employing an efficient algorithm for detection.

\subsubsection{System's Weakness} The fundamental principle of CL systems is to compensate for the errors calculated by the sensor feedback~\cite{nise2020control}. By exploiting the compensation mechanism, compromised sensor feedback can lead to severe system consequences~\cite{tu2019trick, barua2020hall}. For example, a stabilizer uses feedback from inertial sensors to correct its orientation. If this feedback is faulty, the stabilizer may overcompensate, effectively turning into a destabilizer or shaker~\cite{tu2018injected, ji2021poltergeist, zhu2023tpatch}. This weakness is particularly critical when the sensor measures first-order or even multiple-order derivatives of the system's state, e.g., the gyroscope measures angular velocity, which is the first-order derivative of heading angle. In such cases, the errors induced by compromised sensors accumulate over time, causing system-level errors to grow exponentially~\cite{shen2020drift, quinonez2020savior}.

\subsection{Implications on Multi-Sensor Fusion}

For advanced, especially safety-critical applications, CPS equips multiple sensors, which can be either homogeneous or heterogeneous, to perceive the same target comprehensively~\cite{wang2019multi}. The measurements from different sensors are collected by fusion algorithms, e.g., a Kalman filter~\cite{kalman1960new}, to achieve a refined estimate, as illustrated in Fig.~\ref{fig:system}. Since most sensor security studies focus on single sensors, it is crucial to understand their implications on multi-sensor fusion (MSF) systems.

\subsubsection{System's Strength} The redundancy of sensors in MSF systems enables cross-verification, which is commonly believed to be a reliable countermeasure in sensor attack papers~\cite{ji2021poltergeist, tu2019trick, shin2017illusion, xu2018analyzing, sugawara2020light, kohler2022signal}. The ability for sensor data cross-verification is supported by the assumption that a portion of sensors are not compromised, and their data is trustworthy. It is challenging to simultaneously attack multiple sensors even if they are homogeneous, e.g., different types of IMUs usually have different resonant frequencies~\cite{son2015rocking, trippel2017walnut}. To achieve cross-verification, outlier detection and filtering based on statistics is the main technique~\cite{ting2007kalman}. 

\subsubsection{System's Weakness} While it is widely accepted that MSF increases the robustness of sensor measurements, a few studies~\cite{nashimoto2018sensor, shen2020drift} have shown that it is not effective against well-designed sensor spoofing. The basic idea of compromising MSF is to attack the most critical sensor that determines the fusion result. \citet{nashimoto2018sensor} studied attitude-heading reference systems, which estimate the inclination based on the fusion of gyroscope, accelerometer, and magnetometer. It is found that single-sensor spoofing can tamper with the fused results by exploiting the dominant sensor mechanism~\cite{nashimoto2018sensor}. \citet{shen2020drift} found a similar mechanism existed in another fusion algorithm of GPS, LiDAR, and IMU for autonomous driving. They demonstrated that in certain cases where the uncertainties of both IMU and LiDAR measurements are high, the GPS becomes dominant, and then GPS spoofing can take over the fusion algorithm~\cite{shen2020drift}.

\subsection{Implications on Intelligent Perception}

Artificial intelligence (AI) is crucial to extract useful information from those sensors, e.g., cameras, microphones, and LiDARs, whose raw data is highly unstructured~\cite{krizhevsky2012imagenet}. As shown in Fig.~\ref{fig:system}, a typical intelligent perception system involves an AI algorithm, usually a deep neural network (DNN), which recognizes sensor measurements to obtain high-level information, such as classification, detection, and segmentation.

\subsubsection{System's Strength} Conventional sensor attacks~\cite{kune2013ghost, zhang2017dolphinattack, yan2016can, petit2015remote, shin2017illusion} may not affect the intelligent perception results effectively, due to the complex and opaque relationship between sensor data and AI recognition. For example, the replaying commands injected by ultrasonic sound injection~\cite{zhang2017dolphinattack} cannot easily pass intelligent speaker verification~\cite{li2023enrollment}. Furthermore, AI algorithms are resistant to noisy or limited corruption since they can selectively perceive useful information~\cite{cao2019advpoint, man2020ghostimage, yan2022rolling}, making it difficult to achieve the required level of arbitrary manipulation via signal attacks. 

\subsubsection{System's Weakness} For intelligent perception systems, it is crucial to involve the guidance of AI vulnerability for successful sensor attacks. Manual analysis can reveal explicit weaknesses, like in \cite{sun2020towards}, where tiny point clouds were misclassified by object detection models and exploited via LiDAR spoofing. However, in practice, many systems are black-box and thus their weaknesses are implicit and harder to exploit. A popular methodology considers it as an end-to-end optimization problem, where the attack signals are the optimization variable~\cite{sayles2021invisible, man2020ghostimage, cao2019advpoint, jin2023pla}. The optimization is accomplished by gradient descent, similar to adversarial example attacks~\cite{madry2017towards, carlini2017towards}. When the gradient information is unavailable, the optimization method is alternated by black-box methods like grid search~\cite{yan2022rolling} and Bayesian optimization~\cite{ji2021poltergeist}. Moreover, recent work explored the use of unique corruptions by sensor attacks to trigger neural network backdoors~\cite{zheng2023silent,li2023enrollment} or adversarial patches~\cite{zhu2023tpatch}.

\section{Defense Systematization}\label{sec:defense}
In this section, we provide a systematic analysis of defense strategies against sensor OOB vulnerabilities. Our study compares these strategies across several key dimensions, including attack vector coverage, defense effectiveness, deployment overhead. This framework facilitates a qualitative evaluation of defense mechanisms using normalized metrics. A structured overview of the results is summarized in Table~\ref{tab:defense}.

\newcommand{\layman}{\raisebox{-0.2\height}{\includegraphics[width=0.01\textwidth]{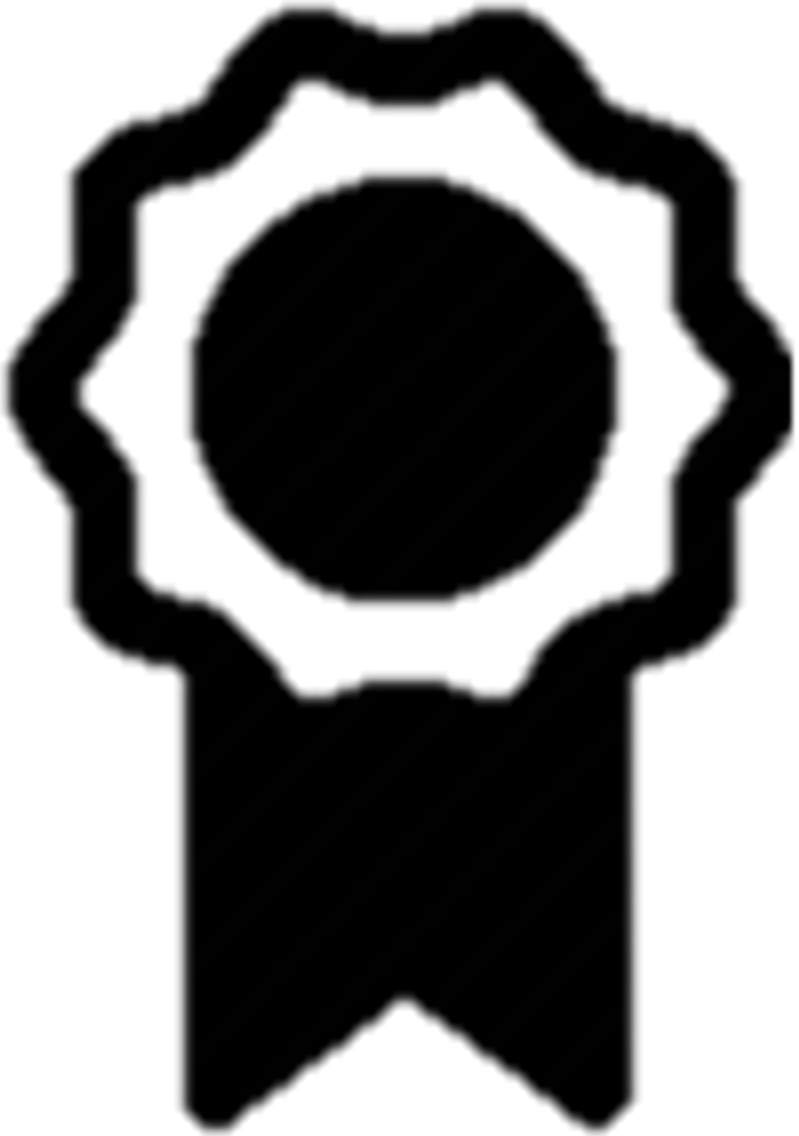}}}
\newcommand{\proficient}{\raisebox{-0.2\height}{\includegraphics[width=0.01\textwidth]{icon/medal.png}\includegraphics[width=0.01\textwidth]{icon/medal.png}}}
\newcommand{\expert}{\raisebox{-0.2\height}{\includegraphics[width=0.01\textwidth]{icon/medal.png}\includegraphics[width=0.01\textwidth]{icon/medal.png}\includegraphics[width=0.01\textwidth]{icon/medal.png}}}
\newcommand{\less}{\raisebox{-0.2\height}{\includegraphics[width=0.025\textwidth]{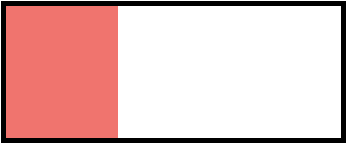}}}
\newcommand{\medium}{\raisebox{-0.2\height}{\includegraphics[width=0.025\textwidth]{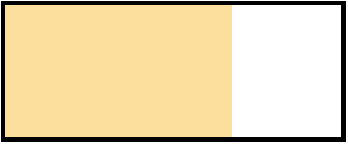}}}
\newcommand{\highly}{\raisebox{-0.2\height}{\includegraphics[width=0.025\textwidth]{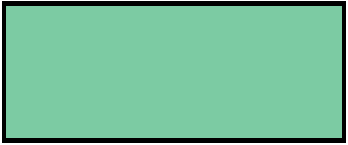}}}
\newcommand{\done}{\raisebox{-0.2\height}{\includegraphics[width=0.015\textwidth]{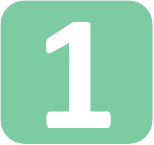}}}
\newcommand{\dtwo}{\raisebox{-0.2\height}{\includegraphics[width=0.015\textwidth]{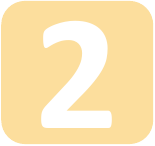}}}
\newcommand{\dthree}{\raisebox{-0.2\height}{\includegraphics[width=0.015\textwidth]{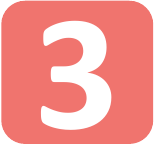}}}

\begin{table*}
    \caption{Systematization of defense methods}
    \resizebox{\textwidth}{!}{
    \begin{tabular}{|c|c|c|c:c:c:c|c:c:c|c:c:c:c|c|}
    \hline
    \multicolumn{2}{|c|}{\multirow{2}{*}{\textbf{Defense Level}}} & \multirow{2}{*}{\textbf{Defense Methods}} & \multicolumn{4}{c|}{\textbf{Targeted Attack Type}} & \multicolumn{3}{c|}{\textbf{Defense Goal}} & \multicolumn{4}{c|}{\textbf{Overhead}} & \multirow{2}{*}{\textbf{Paper}} \\ \cline{4-14}
    \multicolumn{2}{|c|}{} &  & \makebox[0.02\textwidth]{\faVolumeUp} & \makebox[0.02\textwidth]{\faSun} & \makebox[0.02\textwidth]{\faWifi} & \makebox[0.02\textwidth]{\faBolt} & \makebox[0.02\textwidth]{\textbf{C}} & \makebox[0.02\textwidth]{\textbf{I}} & \makebox[0.02\textwidth]{\textbf{A}} & \makebox[0.02\textwidth]{Exp.} & \makebox[0.02\textwidth]{Cost} & \makebox[0.02\textwidth]{Usab.} & \makebox[0.02\textwidth]{Maint.} &  \\ \hline
    \multirow{7}{*}{\rotatebox{90}{\makecell{\textbf{Component}\\ \textbf{Level}}}} & Trans. & Transducer response limitation & \C & \C & \C & \c & \less & \highly & \highly & \proficient & \done & \done & \dtwo & \cite{zhang2017dolphinattack, yan2019feasibility} \\ \cline{2-15}
     & \multirow{4}{*}{\makecell{Signal \\ Cond. \\Circuit}} & Filter stopband enhancement & \C & \C & \C & \C & \less & \highly & \highly & \expert & \dtwo & \dtwo & \dthree & \cite{trippel2017walnut, ji2021poltergeist, son2015rocking, tu2018injected, tu2019trick} \\ \cline{3-15}
     &  & AMP nonlinearity reduction & \C & \C & \C & \C & \less & \highly & \medium & \expert & \dtwo & \done & \dtwo & \cite{zhang2017dolphinattack,yan2019feasibility, roy2018inaudible, trippel2017walnut} \\ \cline{3-15}
     &  & ADC anti-aliasing & \C & \C & \C & \C & \less & \highly & \highly & \proficient & \done & \done & \done & \cite{michalevsky2014gyrophone} \\ \cline{3-15}
     &  & CMRR enhancement & \c & \c & \C & \C & \less & \medium & \medium & \expert & \dtwo & \done & \dthree & \cite{xu2021inaudible, jiang2022wight, jiang2023marionette, jiang2024powerradio} \\ \cline{2-15}
     & Power & PSRR enhancement & \c & \c & \C & \C & \less & \medium & \medium & \proficient & \done & \done & \dtwo & \cite{wang2023volttack, jiang2025vphanton} \\ \cline{2-15}
     & Comm. & Crosstalk noise isolation & \c & \c & \C & \C & \less & \medium & \medium & \expert & \dthree & \dtwo & \dthree & \cite{wang2023volttack, jiang2025vphanton,cronin2021charger} \\ \hline
    \multirow{3}{*}{\rotatebox{90}{\makecell{\textbf{Sensor}\\ \textbf{Level}}}} & Input & OOB signal shielding & \C & \C & \C & \C & \highly & \highly & \highly & \proficient & \dtwo & \dtwo & \dtwo & \cite{yan2020surfingattack, ji2021poltergeist} \\ \cline{2-15}
     & \multirow{2}{*}{Output} & Randomization & \C & \C & \C & \C & \medium & \highly & \less & \proficient & \done & \dtwo & \done & \cite{trippel2017walnut, yan2022rolling, jin2023pla, gluck2020spoofing, fernandez2024deep} \\ \cline{3-15}
     &  & Data encryption & \C & \C & \C & \C & \highly & \less & \less & \proficient & \dtwo & \dtwo & \dtwo & \cite{zhang2017security} \\ \hline
    \multirow{5}{*}{\rotatebox{90}{\makecell{\textbf{System}\\ \textbf{Level}}}} & H. & Redundancy deployment & \C & \C & \C & \c & \less & \medium & \highly & \proficient & \dthree & \dthree & \dthree & \cite{jin2023pla, zhu2023tpatch, shin2017illusion,kohler2022signal} \\ \cline{2-15}
     & H.\&S. & Anomaly detection & \C & \C & \C & \C & \less & \highly & \medium & \expert & \dtwo & \dtwo & \dthree & \cite{tu2021transduction, yan2020surfingattack, yan2019feasibility, park2016ain, tu2019trick} \\ \cline{2-15}
     & \multirow{3}{*}{S.} & Access authentication & \c & \C & \c & \C & \highly & \medium & \medium & \layman & \done & \dtwo & \done & \cite{jiang2022wight, jiang2023marionette, jin2023pla} \\ \cline{3-15}
     &  & Model fusion & \C & \C & \C & \C & \less & \highly & \medium & \expert & \dthree & \dthree & \dtwo & \cite{jin2024unity,bhupathiraju2023emi,cao2023you} \\ \cline{3-15}
     &  & Data recovery & \C & \C & \C & \C & \less & \medium & \highly & \expert & \dtwo & \done & \dtwo & \cite{ji2021poltergeist, kune2013ghost} \\ \hline
    \end{tabular}
    }
    \begin{tablenotes}
        \item H. Hardware \quad H.\&S. Hardware or Software \quad S. Software \quad \C Applicable \quad \c Not applicable \quad \textbf{C} Confidentiality \quad \textbf{I} Integrity \quad \textbf{A} Availability
        \item \less\ Less effective \medium\ Moderately effective \highly\ Highly effective \quad \done\ Low \dtwo\ Medium \dthree\ High \quad \layman\ Layman \proficient\ Proficient \expert\ Expert   \end{tablenotes}
    \label{tab:defense}
\end{table*}

\renewcommand{\layman}{\raisebox{-0.1\height}{\includegraphics[width=0.01\textwidth]{icon/medal.png}}}
\renewcommand{\proficient}{\raisebox{-0.1\height}{\includegraphics[width=0.01\textwidth]{icon/medal.png}\includegraphics[width=0.01\textwidth]{icon/medal.png}}}
\renewcommand{\expert}{\raisebox{-0.1\height}{\includegraphics[width=0.01\textwidth]{icon/medal.png}\includegraphics[width=0.01\textwidth]{icon/medal.png}\includegraphics[width=0.01\textwidth]{icon/medal.png}}}
\renewcommand{\less}{\raisebox{-0.1\height}{\includegraphics[width=0.025\textwidth]{icon/red.pdf}}}
\renewcommand{\medium}{\raisebox{-0.1\height}{\includegraphics[width=0.025\textwidth]{icon/orange.pdf}}}
\renewcommand{\highly}{\raisebox{-0.1\height}{\includegraphics[width=0.025\textwidth]{icon/green.pdf}}}
\renewcommand{\done}{\raisebox{-0.1\height}{\includegraphics[width=0.018\textwidth]{icon/1-2.pdf}}}
\renewcommand{\dtwo}{\raisebox{-0.1\height}{\includegraphics[width=0.018\textwidth]{icon/2-2.pdf}}}
\renewcommand{\dthree}{\raisebox{-0.1\height}{\includegraphics[width=0.018\textwidth]{icon/3-2.pdf}}}

\subsection{Systematization Methodology}\label{sec:defense_methodology}
We systematize existing countermeasures based on the system components they modify and categorize them into three levels: 
\textbf{a) Component-level defense:} This category focuses on hardening critical internal modules within sensors to eliminate exploitable vulnerabilities. Key measures include reinforcing the transducer's anti-interference capabilities, optimizing signal conditioning circuits and power supply modules, and mitigating crosstalk among communication interfaces. These enhancements aim to prevent attackers from reshaping signals or physically damaging components, thereby ensuring the intrinsic robustness of core subsystems.
\textbf{b) Sensor-level defense:} These defenses secure the sensor as a whole, protecting both its input and output while maintaining its core functionality. For input protection, techniques like shielding can block out-of-band interference. For output protection, methods such as real-time signal randomization or data encryption can prevent spoofing and information leakage. \textbf{c) System-level defense:} These methods enhance sensor security from a broader system perspective by integrating cross-layer solutions across hardware, software, or both. Hardware-based approaches may include deploying redundant sensors to increase system resilience. In some cases, combined hardware and software solutions are necessary, such as integrating additional devices with detection algorithms to counter sophisticated attacks. Purely software-based methods often employ AI-driven techniques for tasks like access control, model fusion, or data recovery. This multi-dimensional framework strengthens security at the cyber-physical system level, enabling coordinated defense beyond the limitations of individual sensors.

To facilitate a comprehensive analysis of sensor defense methods, we compare each approach across three key dimensions: targeted attack types, defense goals, and deployment overhead. 

\subsubsection{Target Attack Types} We classify the targeted sensor attacks by their signal modality: attacks by sound/ultrasound (\faVolumeUp), attacks by lasers (\faSun), attacks by radiated EMI (\faWifi), and attacks by conducted EMI (\faBolt). Some defenses are broadly applicable, while others are modality-specific. For example, sensor redundancy is ineffective against conducted EMI attacks, as the injected signal can propagate to all redundant sensors.

\subsubsection{Defense Goal} We use the classical CIA (Confidentiality, Integrity, Availability) security model to assess each defense's primary protection focus, and the effectiveness is categorized into three levels: low (\less), moderate (\medium), and high (\highly). Confidentiality (\textbf{C}) means preventing unauthorized access or inference of sensitive sensor data. Integrity (\textbf{I}) indicates defenses can ensure sensor readings are not altered or spoofed. Availability (\textbf{A}) means maintaining reliable sensor operation despite external disruptions. 

\subsubsection{Overhead} We evaluate real-world feasibility based on four sub-factors: Expertise Knowledge represents the required technical skill level for defense implementation, and we divide it into three levels: layman (\layman), proficient (\proficient), expert(\expert)). \layman \ means the defense is easy to apply and only requires little or no technical background. \proficient\ means the defense needs basic technical skills, and is suitable for trained staff. \expert \ means the defense method requiring deep technical knowledge of both the hardware and software of sensors. 
We rate the remaining three overhead factors as Low(\done), Medium(\dtwo), and High (\dthree). Deployment Cost refers to financial and hardware resource demands. Usability indicates the impact on user experience or system functionality. Maintenance assesses ease of upkeep and adaptability to evolving threats. 

\subsection{Review of Existing Work}
\subsubsection{Component-level Defense}\label{sec:defense_component}
\textbf{a) Transducer.} Limiting the transducer's response range can reduce its susceptibility to abnormal stimuli, thereby mitigating the risk of malicious signal injection. For example, prior studies~\cite{zhang2017dolphinattack, yan2019feasibility} suggest that microphones should be designed to be insensitive to ultrasonic waves. \textbf{b) Signal Conditioning Circuit.} Enhancing the performance of signal conditioning circuits is an effective defense strategy. Key methods include improving the filter's stopband to suppress unwanted frequencies~\cite{trippel2017walnut, ji2021poltergeist, son2015rocking, tu2018injected, tu2019trick}, increasing the amplifier's linearity to minimize distortion and the generation of new frequency components~\cite{zhang2017dolphinattack, yan2019feasibility, roy2018inaudible, trippel2017walnut}, applying anti-aliasing techniques at the analog-to-digital converter to prevent signal misinterpretation~\cite{michalevsky2014gyrophone}, and enhancing the common-mode rejection ratio (CMRR) to resist differential noise~\cite{xu2021inaudible, jiang2022wight, jiang2023marionette, jiang2024powerradio}. 
\textbf{c) Power Supply.} Enhancing the power supply rejection ratio (PSRR) helps shield internal circuits from adversarial power fluctuations. Wang et al.~\cite{wang2023volttack} demonstrate that PSRR can be improved by modifying the architecture of the low-dropout regulator. \textbf{d) Communication Interface.} Isolating crosstalk noise preserves data integrity and prevents unintended signal propagation between channels, as shown in recent work~\cite{wang2023volttack, jiang2025vphanton,cronin2021charger}.

\subsubsection{Sensor-level Defense}\label{sec:defense_sensor}
At the sensor level, defense mechanisms aim to protect the sensor as a unified entity by securing its \textbf{input} and \textbf{output} pathways. 
\textbf{a) Input protection.} Implementing \textit{out-of-band (OOB) signal shielding} is an effective strategy. This approach involves using physical or electromagnetic shielding materials to block unintended or malicious signals such as ultrasound~\cite{yan2020surfingattack, ji2021poltergeist}, light~\cite{sugawara2020light} or electromagnetic waves~\cite{dayanikli2020electromagnetic, wang2022ghosttouch} from reaching the sensor's input interface or being leaked from the sensors~\cite{long2024emeye}, thereby preserving signal integrity and confidentiality. \textbf{b) Output protection.} On the output side, two key techniques are commonly employed. \textit{Randomization} involves introducing controlled variability into the sensor output, such as frequency hopping~\cite{gluck2020spoofing}, randomized sampling~\cite{trippel2017walnut}, rolling shutter sequence~\cite{yan2022rolling},  pulse~\cite{jin2023pla} or pixel noises~\cite{fernandez2024deep}, to make it more difficult for attackers to infer or replicate true measurements. This method can disrupt attempts at sensor spoofing or replay attacks. Meanwhile, \textit{data encryption} secures the sensor's output during transmission by encoding the data, ensuring that even if intercepted, it cannot be easily interpreted or altered~\cite{zhang2017security}. These output-level methods are especially crucial in networked or distributed sensor systems, where data confidentiality and authenticity are paramount. Together, these input and output protections strengthen the end-to-end security of the sensor against a wide range of physical-layer and signal-based attacks.

\subsubsection{System-level Defense}\label{sec:defense_system} 
System-level defense mechanisms aim to improve overall system resilience using both hardware and software rather than protecting individual sensor elements alone. 
\textbf{a) Hardware-based defenses.} One widely used strategy is the deployment of redundant sensors, where multiple sensors of the same or different types are used to measure multiple physical quantities that are related to the same task, such as combining lidars and cameras in automotive systems~\cite{jin2023pla, zhu2023tpatch, shin2017illusion,kohler2022signal}, enabling error correction or failover mechanisms. \textbf{b) Hardware\&Software-based defenses.} Hybrid methods like anomaly detection~\cite{tu2021transduction, yan2020surfingattack, zhang2017dolphinattack,yan2019feasibility, roy2018inaudible, zhu2023tpatch, davidson2016controlling, park2016ain, dai2023inducing, tu2019trick} are commonly applied to identify unexpected behaviors or outputs that may indicate tampering or spoofing. For instance, study~\cite{tu2021transduction} suggests using a matched dummy sensor circuit that shares the sensor’s vulnerabilities to EMI but is insensitive to legitimate signals that the sensor is intended to measure to detect sensor attacks. \textbf{c) Software-based defenses.} Access authentication ensures that only authorized entities~\cite{jiang2022wight, jiang2023marionette} can read or write sensor data, or only verified signals~\cite{jin2023pla} can be received and processed, preventing unauthorized control or data leakage; Model fusion~\cite{jin2024unity,bhupathiraju2023emi,cao2023you} integrates data from multiple models or sensor sources to validate and reinforce output correctness, improving robustness against targeted attacks; And data recovery techniques~\cite{ji2021poltergeist, kune2013ghost} attempt to restore accurate sensor outputs from corrupted or missing data, using interpolation, signal reconstruction, or AI-based methods. Together, these system-level methods provide layered protection and enable coordinated threat response across the entire sensor ecosystem.

\subsection{Research Gaps and Future Directions}
Based on the above analysis, we identify key limitations and provide future research directions. 
\begin{itemize}[leftmargin=10pt]
    \item \underline{\textit{Cross-component interactions.}} 
    Most component-level defenses are designed for isolated modules (e.g., transducers, amplifiers), overlooking cross-component interactions that can lead to OOB vulnerabilities. To mitigate this issue, a promising direction is to incorporate safety simulations during the sensor design phase. For instance, researchers can conduct multiphysics simulations~\cite{frangi2008advances} to capture field coupling effects in MEMS transducers and proactively identify resonant acoustic frequencies that may compromise sensor integrity. Similarly, combining electromagnetic field simulations with circuit-level modeling of signal conditioning stages can help reveal electrical characteristics that contribute to OOB vulnerabilities. These approach can inform design decisions early in the development cycle, reducing the risk of post-deployment security issues.
    \item \underline{\textit{Lack of defenses for snooping attacks.}} 
    We find that there are few methods specifically aimed at protecting the confidentiality of sensor measurements. We believe a viable direction is to leverage internal hardware components within the sensor itself for encryption, rather than relying on software-based encryption. For example, prior work has demonstrated that the physical unclonable function (PUF) properties of image sensor phototransistors can be used to encrypt captured signals, enabling verification of authenticity and originality~\cite{shao2023highly}. This in-sensor encryption paradigm can offer a novel approach for protecting privacy-sensitive measurements at the hardware level.
    \item \underline{\textit{High integration complexity of redundancy methods.}}
    Redundancy-based defenses (e.g., sensor fusion) often increase system-level complexity and cost, limiting their practicality in cost-sensitive applications. For example, adding redundant IMUs to a UAV requires changes to both PCB layout and flight control algorithms. Inspired by paper~\cite{zhang2023adc}, which uses redundant ADCs for attack detection, a more lightweight alternative is to embed redundancy components within a single sensor. For example, multiple transducers with distinct resonant frequencies can be integrated into a single accelerometer to ensure at least one remains unaffected during an attack. 
\end{itemize}

\section{Discussion}

\subsection{Comparison with Previous Work}
Compared with previous SoK papers and surveys in this field, our paper differs in the following aspects.

\textbf{Terminology.} 
The term out-of-band has been used inconsistently across prior works. SoK~\cite{yan2020sok} and paper~\cite{barua2022sensor} interpret band as the frequency range of the signal, while paper~\cite{giechaskiel2019taxonomy} adopt definitions from network communication, referring to signals injected through covert communication channels. This inconsistency may lead to confusion and hinder a unified understanding of the field. In this work, we offer a formal and structured definition of OOB vulnerabilities based on physical energy conversion principles. This not only clarifies the boundary between in-band and out-of-band but also lays a conceptual foundation for future research.

\textbf{Scope.} Our SoK centers on the concept of sensor OOB vulnerabilities. In contrast, SoK~\cite{yan2020sok} focuses on transduction attacks and signal injection techniques, without identifying the underlying commonalities in sensor vulnerabilities and covering side-channel and privacy snooping attacks. SoK~\cite{walker2021sok, xu2023sok} focus on specific attack scenarios, i.e., eavesdropping via mobile sensors and sensor spoofing against robotic vehicles.

\textbf{Systematization Methodology.} We for the first time adopt a bottom-up methodology that examines vulnerabilities at the component, sensor, and system levels. In particular, our sensor-level analysis provides a dedicated analysis of the practicality of sensor attacks, which is overlooked in SoK~\cite{yan2020sok}. Our system-level analysis provides new insights into how CPS architecture influence vulnerability exposure and mitigation trade-offs, which is not covered in previous work.

\subsection{Sensor Design Trade-offs}
Sensor design inherently involves trade-offs between performance, cost, and resilience to attack signals. Enhancing sensitivity improves signal detection in low-power applications, but also lowers the threshold for OOB interference. Filtering and shielding can block malicious input, but risk degrading legitimate functionality or increasing size and cost. Software-based defenses (e.g., anomaly detection, sensor fusion) offer flexibility but strain the limited compute and power budgets of embedded CPS. Designing sensors that balance utility with OOB robustness remains an open challenge, especially as sensors become more intelligent and integrated. 
Nonetheless, we encourage sensor designers to explicitly consider OOB vulnerabilities during the design process. For legacy sensors where redesign is impractical, these vulnerabilities should at least be acknowledged in the datasheet, which can provide a primary reference for system developers to raise awareness and guide secure integration.

\subsection{OOB Vulnerabilities for Good}
In this paper, we assume by default that sensors are used for benign purposes. However, sensors can be used for malicious purposes, such as hidden cameras and eavesdropping voice recorders that invade privacy. In such cases, the sensor OOB vulnerabilities can be leveraged to implement proactive defenses. For instance, defenders can detect malicious devices by sensors' EM radiation~\cite{ramesh2022ticktock, chaman2018ghostbuster, liu2023camradar,zhou2023dehirec}. \citet{ramesh2022ticktock} proposed a system to detect microphone status by probing EM emanations from laptop circuitry carrying mic clock signals, which only appear during recording. Similarly, \citet{chaman2018ghostbuster} introduced Ghostbuster, which detects hidden eavesdroppers like cameras by probing RF clock leakage without modifying current transmitters and receivers. Additionally, \citet{liu2023camradar} presented a method to detect hidden cameras by observing changes in EM emanations caused by the camera's clock modulation.

\subsection{Security Outlook for Smart Sensors}
Sensors are increasingly armed with intelligent technologies. In the Internet of Things (IoT) applications, transmitting large volumes of raw sensory data consumes significant communication and computational resources. To alleviate the issue, computational tasks start to be integrated into the sensors, leading to the rise of smart sensors~\cite{zhou2020near, warden2023machine}. 
Nowadays, sensor manufacturers are embarking on the smart sensor market, e.g., voice-wakeup microphones~\cite{knowles}, glass-break sensors~\cite{ring}, and fitness-tracking motion sensors~\cite{bosch}. These smart sensors offer advantages like reducing device power consumption and processing sensitive data within the device~\cite{iot-sensor}, thus minimizing the risk of privacy leakage~\cite{ji2024sensor}.

However, smart sensors are increasingly susceptible to attacks due to their reliance on intelligent algorithms and sophisticated hardware. As discussed in Sec.~\ref{sec:sys}, intelligent perception is vulnerable to targeted attacks that exploit these algorithmic weaknesses~\cite{carlini2017towards}. In addition, the computing hardware within smart sensors is exposed to physical side-channel attacks. Besides sensor readings, these attacks can extract sensitive information, such as model architecture and parameters, by analyzing EM emissions~\cite{batina2019csi, hu2020deepsniffer}.

\section{Conclusion}
    In this SoK, we systematically analyze sensor OOB vulnerabilities from the perspective of energy conversion. By classifying attack signals into \textit{out-of-range} and \textit{cross-field} types and analyzing their propagation paths, we build a unified framework to better understand and formalize sensor attacks. This helps in creating effective detecting and defenses on sensor OOB vulnerability and offers guidance for designing more secure sensors in the future. The study also highlights the wider impact of sensor security on CPSs, calling attention to both the growing range of sensor attacks and the core physical principles behind sensor function.

\section*{Acknowledgment}
We thank the anonymous reviewers for their valuable comments. This work is supported by the National Science Foundation of China (NSFC) Grant 62222114 and 62201503.





\balance
\bibliographystyle{IEEEtranN}
\bibliography{reference/reference.bib}
%

\end{document}